\documentclass[aps,amsmath,twocolumn,amssymb,floatfixng,showpacs,
superscriptaddress,footinbib]{revtex4-1}
\usepackage[T1]{fontenc}
\usepackage{bbm}
\pdfoutput=1
\usepackage[dvips]{graphics}
\usepackage{bm}
\usepackage{float}
\usepackage{epsfig}
\usepackage{enumerate}
\usepackage{subfigure}
\usepackage{amsmath}
\usepackage{color}
\usepackage{braket}
\usepackage{graphicx}
\usepackage{float}
\usepackage[colorlinks=true,linktoc=page,linkcolor=red,citecolor=blue,urlcolor=magenta]{hyperref}

\newcommand{\myvec}[1]{\textbf{#1}}

\newcommand\bea{\begin{eqnarray}}
\newcommand\eea{\end{eqnarray}}
\newcommand\beq{\begin{equation}}  
\newcommand\eeq{\end{equation}}

\begin{document}
\title{Extended Haldane model- a modern gateway to topological insulators}
\author{Tanay Nag}\email{tanay.nag@hyderabad.bits-pilani.ac.in}\affiliation{Department of Physics, BITS Pilani-Hyderabad Campus, Telangana 500078, India} 
\thanks{Tanay Nag, corresponding author}
\author{Saptarshi Mandal}\email{saptarshi@iopb.res.in}\affiliation{Institute of Physics, Bhubaneswar- 751005, Odhisa, India}
\affiliation{Homi Bhabha National Institute, Mumbai - 400 094, Maharashtra, India}
	
\begin{abstract}

The seminal Haldane model brings up a paradigm beyond the quantum Hall effect to look for a plethora of topological phases in the honeycomb and other lattices. Here we dwell into this model considering a full parameter space in the presence of spin-orbit interaction as well as Zeeman field such that the flavour of Kane-Mele model is invoked. Adopting this extended Haldane model as an example, we elucidate, in a transparent manner, a number of topological features in a pedagogical manner. First, we describe various  first order topological insulator phases and their characterizations while explaining various anomalous quantum Hall effects and quantum spin Hall effects in the extended Haldane model.  Second, we demonstrate the concepts of higher order topological insulator phases along with the topological invariants in the 
anisotropic limit of the extended Haldane model.  At the end, we discuss various open issues involving \textcolor{black}{emergent or extended} symmetries that might lead to a broader understanding of various topological phases and the associated criteria behind their emergence.   
\end{abstract}

	
\date{\today}
	
\maketitle
\section{Introduction}
\label{intro}
Recently the study   of topological aspects of various condensed matter systems has established itself as a separate field of research. Historically topology is a mathematical concept that examines the shapes of the object under continuous deformations without puncturing it and ascribes an integer or some appropriate index to identify the topological class of an object that it belongs to \cite{millman-parker,mukhi-mukunda,nakahara-2018,spivak-1999,engelking-1989}. The two seemingly different objects can be shown to have the same topological class if they can be mapped to each other under a continuous deformation which often is described as adiabatic transformation in physics community. For condensed matter systems, recently,  these studies gained wide attention due to many reasons. First is to find examples within a  model building approach of new topological states of matter\cite{shun-qing-2013,bernevig-2013,asboth-2016}. Secondly, more interesting and challenging aspects are examined to design suitable experiments to detect such topological phases by appropriate transport measurement\cite{ying-xing,hua-ding,nini-2019}. Lastly, various synthetic platforms such as acoustic \cite{ma-chan-nature-review} and photonic circuits\cite{ozawa-2019-rmp,zhihao-2022} are being used to mimic the quantum mechanical systems to realize the long-lived edge states which are Hallmark of topological insulators. The role of interaction, disorder and its interplay with various symmetries are the future directions of studies in topological insulators\cite{binglan,prodan-2011,hormann}. As the topological states of matter are robust against small to moderate perturbations, they promise many useful applications for various practical uses\cite{tian-mdpi,kushal-2021,you-2020,wang-2019}. \textcolor{black}{Owing to the energy efficiency and topological protection of the edge modes in such topological materials, they can have potential applications in the field of topological electronics and sensors that can outperform the Silicon-based technology in the near future \cite{gilbert21, liu20}.}\\

\indent
Experimental discovery of superconductivity in 1911 by \cite{kamerlingh} Heike Kamerlingh Onnes brought into knowledge of  an unknown state of matter which culminated in celebrated BCS theory\cite{bcs-1957} establishing a paradigm of itself based on Landau theory phase transition. Going beyond the realm of Landau theory of phase transition,  another paradigm of research namely, the topological phases of matter emerges.  
 The prototypical experimental discovery of integer quantum Hall effect (IQHE) \cite{klitzing-1980,klitzing-rmp-1986} revealed an incompressible quantum liquid state in two dimensions with insulating phase in the bulk but conducting channel in the edges. Later the quantized transverse Hall conductivity was shown to be determined by TKNN integer\cite{tknn-1982} which is a topological index known as Chern number. This initiates the journey of another field of research that is aptly described today as the  topological Chern insulator. Soon after the Nobel prize was conferred to Klaus von Klitzing in the year 1985 for the discovery of IQHE, F.D.M. Haldane introduced a tight binding model\cite{haldane-1988} on graphene with no net magnetic field but showing a quantized Hall conductivity. The resulting quantum Hall state is known as the quantum anomalous Hall effect (QAHE). The implications of this seminal paper were revived when Kane-Mele proposed a spin-full version of Haldane model\cite{kane-2005-1st,kane-2005-2nd} with spin-orbit coupling (SOC) and which gave birth to a new state of matter known as quantum spin Hall effect (QSHE). \textcolor{black}{The SOC  and  the magnetic field induce intriguing effects on the band dispersion, respectively, for the time reversal symmetry (TRS) preserved QSHE and TRS broken IQHE, leading to the topological gap. }\\

\indent
One may note that the QAHE is realized in a time reversal broken system and is characterized by the topological index  known as first Chern number. The QSHE respects TRS and thus Chern number no longer remains relevant as a meaningful topological characterization. Instead, the QSHE is described by what is known as $Z_2$ index and gives birth to a new class of topological phases referred to as $Z_2$ topological insulator. Interestingly, the first Chern no depends on the Berry connection $\vec{A}_n(k_x,k_y)= \langle u_{n,k}| \vec{\nabla} |u_{n,k}\rangle $\cite{berryphase-1984}  and defined as $C_n = \int F_n(k_x, k_y) dk_x dk_y$ with $\vec{F}_n(k_x, k_y)= \vec{\nabla} \times \vec{A}_n(k_x, k_y)$. Here `$n$' denotes the band index and $|u_{n,k}\rangle$ denotes an eigenstates of the $n$-th band. Due to  certain symmetry of $\vec{F}_n(k_x,k_y)$, $C_n$ vanishes in the presence of TRS. On the contrary  the $Z_2$ index \cite{kane-2005-1st,kane-2005-2nd,soluyanov-2011} relies on partitioning the momentum space into even and odd subspaces such that $\Theta H(k) \Theta^{-1} =   H(-k)$ and $\Theta H(k) \Theta^{-1} \ne  H(-k)$ for even and odd subspaces, respectively, where $\Theta$ is the operator for TRS. For the even subspace the wavefunction $|u(k)\rangle$ and $\Theta |u(k) \rangle$ are connected by a $U(2)$ transformation whereas in the odd subspace $|u(k)\rangle$ and $\Theta |u(k) \rangle$ are orthogonal. The $Z_2$ index is obtained by evaluating the winding of the phase of $P(k)= {\rm Pf} [\langle u_i(k)|\Theta|u_j(k)\rangle]$ around a loop enclosing half of the Brillouin zone (BZ). In fact, this is a reminiscence of  Chern number that depends on the number of patches where wavefunctions are not smoothly connected by a multiplication of simple phase factor and thus  counts the number of such patches and is determined by the phase accumulated at the boundary of such patches. This suggests that there might be an implicit connection between these two topological characterizations and spin-Chern number\cite{prodan-2009,prodan-2010,spin-chern1} made this connection more evident. The present review would explicitly use the spin-Chern number to characterize the topological phases and show the role of a finite spin-spectrum gap. The connection between $Z_2$ index and spin-Chern number presents an intriguing addition to the classification of topological phases based on symmetries \cite{10fold,kitaev-2009}and may lead to deeper understanding for further applications in the presence of additional emergent\textcolor{black}{/extended} symmetries and higher dimensions\cite{andreas-3d-2008,rahul-prb-2009,rahul-njp-2010,hassan-3d-anrev}. \textcolor{black}{The spin-Chern number happens to be a good topological marker irrespective of the TRS of the problem as long as there  exists a gap in the bulk dispersion. }\\

\indent
All the above mentioned topological insulators are characterized by a bulk insulating gap in $D$ dimensions with a conducting edge channel in $D-1$ dimensions referring to first order topological insulator (FOTI). Note that for $D=3$ similar extensions of the $Z_2$ topological insulators are realized with gapless surface states \cite{rahul-prb-2009,rahul-njp-2010,fu-kane-mele-2007,rahul-z2-2009}. In a concomitant development it was realized there exists another class of topological insulators for which the boundary states exist in $D-n$ dimensions with $n \ge 2$ \cite{schindler-2018,benalcazar-2017,benalcazar-2017-prb,parames-2017}. These are dubbed as higher order topological insulator (HOTI) and currently a very thriving area of research. It is realized that it is essential to break the point group symmetry $C_4$ or $C_3$ for square and honeycomb lattice respectively (by introducing certain anisotropy) to realize such HOTI phases. In these phases the bulk system is characterized by the macroscopic charge polarizations or various higher order charge moments such as dipole, quadrupole and octupole moments \cite{benalcazar-2017,benalcazar-2017-prb}. The fundamental difference between the FOTI phase and HOTI phase is that  in the former phase  the system is characterized by extended wave-function given by Bloch bands while in the later phase, the extended wavefunction as obtained from Bloch band can not be used directly to account for the localized charge polarizations \cite{resta-1992-1,resta-1994-2,resta-1998-3} hence Wannier functions are used. Interestingly underlying mathematical formulation based on the adiabatic evolution connects the first order and second/higher order topological insulator through Berry potential \cite{resta-1994-2}. Under specific symmetries, the higher order moments such as dipole moment, quadrupolar moment and octupolar moments are quantized and serve as the relevant topological order parameters \cite{benalcazar-2017,benalcazar-2017-prb}. \textcolor{black}{In a nutshell, the field of higher order topological systems extends framework of topological invariant to a generalized platform that also includes first-order characterizations. }\\

\indent
The purpose of this review is to present the salient aspect of recent developments of first order and second order topological insulator through the celebrated Haldane model, followed by Kane-Mele model and  detailed analysis of an extended Haldane model  that bridges Haldane model\cite{haldane-1988} and Kane-Mele model \cite{kane-2005-1st,kane-2005-2nd} giving rise to the all possible FOTI phases that can appear. Then as an example of a model for HOTI we present the second order topological insulator (SOTI) phase that could arise within the extended Haldane model by introducing certain anisotropy. We discuss all the technicalities of the model, the role of various symmetries and  its relation to the appearances of edge states in FOTI and SOTI along with the possible future directions.  Before defining the model and describing its analysis we first briefly review the literature which has been extensively looking at various aspects of topological insulators taking inspiration from Haldane model, Kane-Mele model and other seminal models.\\

\indent
As far as the topological aspect of non-interacting tight-binding models are concerned, there are remarkable progress and understanding yielding varieties of quantum Hall systems such as, quantum anomalous Hall insulator (QAHI) \cite{haldane-1988,nagaosa-2003} and quantum spin Hall insulator (QSHI) \cite{kane-2005-1st,kane-2005-2nd,barnevig-2006} and many more \cite{klitzing-rmp-1986,QHE_review2,QHE_review3,QHE_review4}. All of these are categorized by the specific way the edge charge and spin current behave. For example the QAHI\cite{haldane-1988} has quantized charge current whereas  and QSHI \cite{kane-2005-1st} posses quantized  spin current. This distinct nature of the edge current is manifestation of the absence(for QAHI) or presence (for QSHI) of TRS symmetry.  For this reason, the spin-orbit coupling becomes the basic ingredient for QSH effect to be realized.  The bulk topological invariant Chern number \cite{niu-85,kohmoto-85}  can successfully predict the number of edge states in QAHI  phases \cite{bbc,essin-2011,li12}. On the other hand, QSHI phase, characterized by spin-polarized edge currents, is described by an equivalent topological invariant known as spin-Chern number \cite{haldane06,sheng-prl-2005}. The time reversal invariant
QSHI phase can also be equivalently described by a $Z_2$ topological index \cite{kane-2005-1st,prodan-2009} and its relation to spin-Chern number has been well established and can equivalently classify TRS invariant system \cite{spin-chern1,spin-chern2,spin-chern3,spin-chern4}. It may also be noted that mirror symmetry breaking Rashba SOC term does not destroy the topological order of the QSHI state even though the spin conservation no longer holds. \textcolor{black}{We note that there exist other two important symmetries namely, particle-hole and chiral symmetry which play crucial roles in protecting the boundary modes of any topological system in general.} \\

\indent
The first part of the present review will focus mainly fate of QSHI when time reversal breaking has been introduced. It may be noted that such efforts have been realized by introducing exchange field \cite{qiao10,sheng-2011} or magnetic doping  \cite{Liu-2008,li2013}. Staggered magnetic flux \cite{luo-2017} in square lattice are also introduced in QSHI system to obtain  QAHE. This  review makes the connection from QAHI\cite{haldane-1988,kim16,rachel14,exp1,exp2} to QSHI through various quantum anomalous spin Hall insulator (QASHI) phases, invoking a staggered magnetic field in Kane-Mele model \cite{kane-2005-1st}, more explicit. \textcolor{black}{The QASHI represents a TRS broken QSHI (refer to Sec. \ref{sec:FOTI-model}A page 7 for more detailed discussion)} It further demonstrates that for TRS broken system,   $Z_2$ index is not a relevant topological invariant but the spin-Chern number remains a relevant topological invariant where finite spin-spectrum gap determines the topological phase boundary\cite{sheng-2011,spin-gap}. It may be noted for spin-full generalization of the Haldane model that one can in general obtain eight different non-trivial topological phases as shown in \cite{Saha21}. We denote the spin-Chern number for the up and down-spin sector by $\rm{C}_{\uparrow}$ and $\rm{C}_{\downarrow}$ \cite{prodan-2009} respectively.  For the QASHI phases only one of the spin-Chern number is non-zero and hence denoted by $\rm (C_{\uparrow},0$ or $\rm (0,C_{\downarrow})$. On the other hand for the QSHI phases both the spin-Chern number are finite (they may or may not be equal in sign and magnitude) and represented by $\rm (C_{\uparrow}, C_{\downarrow})$.  \textcolor{black}{Our work thus reveals a variety of topological phases where spin and charge degrees of freedom both play significant roles.}


\indent 
The topological insulators are best known for their intriguing bulk-boundary correspondence. The eight-fold quantum Hall states,  realized in an extended Haldane model, would serve as an ideal platform where the charge quantization and spin-polarization of the edge current could have many possibilities.   For example in QASHI phase the edge current is spin-selective, for  QSHI it is spin-polarized and QAHI phases has spin-neutral edge current. Regarding the bulk-boundary correspondence, the HOTI phases \cite{benalcazar-2017-prb,schindler-2018} show more exotic manifestation \cite{haldane-1988,nagaosa-2003,kane-2005-1st,kane-2005-2nd,barnevig-2006,klitzing-rmp-1986,QHE_review2,QHE_review3,QHE_review4}. Apart from the time-reversal symmetry which was an important ingredient to realize various quantum Hall phases, various other symmetries such as, reflection,  inversion, rotational symmetry plays an instrumental role in determining the HOTI phases \cite{Zhida17,Langbehn17,Khalaf18,Miert18,Dumitru19}. For example zero-dimensional electronic corner modes have been obtained for two-dimensional quantum spin-Hall insulator. On the other hand for topological superconductor the zero energy corner modes become Majorana modes. These zero-energy modes are characterized by quadrupole moment which are quantized due to topological origin \cite{Liu18,Wang18,Volpez19,Roy20,Wang20,Wu20,Ghosh21c,Yang21,Ghorashi2020,QWang18,ghosh-22a,ghosh-22b}. This quantized quadrupole moments manifests through particular edge and bulk polarization of charge density and Wannier centers. Though such HOTI phases are first proposed in simple square and cubic lattices  \cite{Franca18,nag19,Seshadri19,Bomantara19,Huang20,Haiping20,Nag21,Ghosh21,Ghosh21b,Zhu21,RXZhang21,Ghosh22a,
Wang20,Wang20,Chen19,Xie19,imhof2018topolectrical}, later HOTI phases are realized in other lattices \cite{Ezawa18,Liu19,Mizoguchi19,Lee20,Xue21,Zangeneh19,Ezawa18b,van2020topological,el2019corner,li2020higher,xue2019acoustic,ni2019observation,Bunney22}. \textcolor{black}{In this review, we extensively discusses FOTI and HOTI phases based on the honeycomb lattice platform. }


\indent 
In an important study \cite{Wang21} continuing the extension of Haldane model to incorporate a $C_3$ symmetry breaking by introducing anisotropic hopping,  higher-order corner modes are shown to emerge (through a gap closing and reopening phase transition) from a FOTI phase  that hosts first-order QAHI phases otherwise. Interestingly, by making use of pseudospin degree of freedom and anisotropic hopping strength, the honeycomb lattice is examined to host helical edge state in a HOTI phase even without SOC interaction  \cite{Liu19}. Such modulation in hopping for the honeycomb lattice can in principle lead to a mismatch between the Wannier centers, defined by the expectation value of the projected position operator on the occupied band, in the unit cell of a crystal and the lattice sites  \cite{benalcazar-2017,Zhida17,benalcazar-2017-prb,Ezawa18}. The WC, being related to $d$-dimensional polarization, is located  at a high-symmetry point with respect to the mirror symmetries and remains quantized for the HOTI phase \cite{resta-1994-2,Smith93}. Notice that the spin-Chern number \cite{prodan-2009,prodan-2010,spin-chern3} and $Z_2$ invariant \cite{kane-2005-2nd}, defined for the FOTI phase, vanish in the SOTI phases being the primary criteria to look for the SOTI phase. In the anisotropic version of Haldane model \cite{Wang21} it was realized that as long as the inversion symmetry (IS) is preserved the dipole moment is quantized to 0.5. This review  considers  TRS broken QSHI \cite{qiao10,sheng-2011,Liu-2008,li2013}, based on graphene and shows the effect of a Zeeman field as well. Notabely an external Zeeman field can also host HOTI phases provided IS is preserved \cite{Yafei20} which could be thought of as an extension of Kane-Mele model(QSHI phase)  in presence of external magnetic field.
Similarly analogous QAHI phases can be produced in the absence of sub-lattice mass \cite{Wang21}. In the second part of this review, we discuss how the bulk-boundary correspondence for the $C_3$ symmetry broken  Haldane model produces many realization of HOTI phases where Zeeman exchange field, SOC interaction and sub-lattice mass interplay with each other in a non-trivial way\cite{saha-23}.

\textcolor{black}{
Before moving to the review of extended Haldane
model and how it encompasses different topological
phases introduced, we now briefly mention the remarkable achievements to realize many such topological phases. The QAH phase has been predicted in Hg$_{1-y}$Mn$_y$Te \cite{Liu-2008}, a family of 2D organic topological
insulators \cite{Wang13a}, graphene \cite{qioa-2014} and magnetically doped
InAs/GaSb \cite{Wang14} and also in other systems \cite{Deng22}.  QAHE has been experimentally confirmed  in many
systems such as chromium-doped (Bi, Sb)$_2$Te$_3$ \cite{Chang13} and
magnetically doped topological insulator grown on the
antiferromagnetic insulator  Cr$_2$O$_3$ \cite{Pan20}. For excellent
review on various challenges in realizing QAHE  and recent progress, we refer to Refs.  \cite{QHE_review4}, \cite{Chi22}, and \cite{He14}.
On the other hand, QSHE was predicted and then observed in HgTe/(Hg,Cd)Te quantum wells \cite{exp2} at some
critical thickness of the 2D sample. Other experimental realization of QSHE includes monolayer tungsten
ditelluride \cite{Tang17,Wu18} (WTe$_2$) and in inverted InAs/GaSb
quantum wells \cite{Kenz11}. In addition to the FOT phases, there have been significant experimental progress in realizing the HOT phases in solid-state systems \cite{Schindler18,Noguchi21} as well as 
meta-materials \cite{Garcia18,Xue19,Weiner19,Imhof18,Zhu20, Zhang19}.  Given the experimental advancement in the field of topological phases, the theoretical understanding discussed in this  review may be helpful in designing future experiments.}


Thus in a nutshell this review considers the dual aspect of topological insulator mainly FOTI and SOTI within the ambient of the extended Haldane model and thus constitutes an important reference of many prevalent notions of topological insulators.

\section{Chern insulator and quantum spin Hall insulator}
\label{sec:CI and QSHI}

Given the bipartite nature of the honeycomb lattice, the graphene can be made a Chern insulator under suitable conditions \cite{Haldane-88,Jotzu14}. The TRS can be broken by the introduction of magnetic flux while IS is broken by an onsite sub-lattice mass term. This causes a gap in the  Dirac spectrum while the topological nature of the gap is 
ascertained when the gap changes sign between two Dirac nodes. In this context, Haldane model is a prime example of QAHI where the topological phase breaks unitary chiral, anti-unitary particle-hole symmetries and TRS residing in class A in the ten-fold classification table. The model Hamiltonian is given by  
\begin{eqnarray}
\label{main:ham_Haldane}
H = && -t_1 \sum_{\langle ij \rangle} c^{\dagger}_i c_j +  t_2 \sum_{\langle\langle ij \rangle\rangle} e^{i \nu_{ij} \phi } c^{\dagger}_i c_{j} + M \sum_i c^{\dagger}_i \zeta_i c_i  \nonumber \\
\end{eqnarray}
where $c_i$ ($c^{\dagger}_i$) represents the fermionic annihilation (creation) operator. The first term represents the regular inter-sub-lattice hopping in graphene. The second term contains the next nearest neighbour hopping in the presence of magnetic flux.
Here, $\nu_{ij}=\pm 1$ refers to the chirality of the magnetic flux $\phi$ contained within the triangle formed by the next nearest neighbor sites i.e., for A (B), it takes the value $+1$ ($-1$). On the other hand, in the third term, $\zeta_i=+1 (-1)$ for A (B) sub-lattices denotes the IS breaking mass term of strength $M$.  The momentum space representation of the model, on the basis $(c^A_{\bm k},c^B_{\bm k})^T$, is given by 
\begin{eqnarray}
H_{\bm k}&=& 2t_2 \cos \phi \sum_{j} \cos ({\bm k}\cdot {\bm a}_j) \sigma_0 \nonumber \\
&+& t_1 \sum_{j} \bigg(  
\cos ({\bm k}\cdot {\bm \delta}_j) \sigma_x + \sin ({\bm k}\cdot {\bm \delta}_j) \sigma_y  \bigg) \nonumber \\
&+& \bigg(M-2 t_2 \sin \phi \sum_{j} \sin ({\bm k}\cdot {\bm a}_j)   \bigg) \sigma_z
\end{eqnarray}
where nearest neighbour vectors ${\bm \delta}_1= a/2(1,-1/\sqrt{3})$, ${\bm \delta}_2= a/\sqrt{3}(0,1)$, ${\bm \delta}_3= a/2(-1,-1/\sqrt{3})$ and next nearest neighbour vectors ${\bm a}_1= {\bm \delta}_1- {\bm \delta}_3 $, and ${\bm a}_2= {\bm \delta}_2- {\bm \delta}_3 $.  We consider $a=1$ for simplicity.

The next nearest neighbour hopping $t_2$, combined with $C_3$ symmetric flux plaquette, leads to an onsite momentum-dependent mass term that competes with the staggered momentum-independent mass term leading to the topological phase boundary $M=\pm 3 \sqrt{3} t_2 \sin \phi$. The $\pm$ sign corresponds to two Dirac points ${\bm K}=(2\pi/3, 2\pi/\sqrt{3})$ or ${\bm K}'=(-2\pi/3, 2\pi/\sqrt{3})$ rendering the fact that the topological mass term $\Delta= \pm M \mp 3 \sqrt{3} t_2 \sin \phi$ changes its sign only for one of the Dirac cones across the phase boundary leaving the sign of the gap for the other Dirac cone unchanged. Therefore,  as discussed earlier, either ${\bm K}$ or ${\bm K}'$ Dirac point is topologically gapped within the topological phase bounded by $|M| < |3 \sqrt{3} t_2 \sin \phi |$. Outside this phase i.e., $|M| > |3 \sqrt{3} t_2 \sin \phi |$, both the Dirac points are trivially gapped with the same sign of  $\Delta$. In order to topologically characterize this quantum Hall phase, one can compute Chern number that changes sign between positive and negative values of $\phi$ when $|M| < |3 \sqrt{3} t_2 \sin \phi |$. On the other hand, for $|M| > |3 \sqrt{3} t_2 \sin \phi |$, Chern number vanishes. This is due to the fact that one wavefunction can cover the whole momentum space Brillouin zone for $|M| > |3 \sqrt{3} t_2 \sin \phi |$ while this no longer remains true for $|M| < |3 \sqrt{3} t_2 \sin \phi |$ due to the sign change in $\Delta$ between two Dirac points.  One can find chiral edge modes for a semi-infinite ribbon geometry as a signature of bulk-boundary correspondence for $|M| < |3 \sqrt{3} t_2 \sin \phi |$.

Very recently, motivated by the Haldane model the Kane-Mele model has been introduced after incorporating the spin degrees of freedom \cite{{kane-2005-1st},kane-2005-2nd}. This model, defined on graphene honeycomb lattice, preserves TRS  and can be thought of as two copies of Haldane model such that the spin up Haldane model $H^{\uparrow}_{\bm k}$ is a time reversal partner of the spin down $H^{\downarrow}_{\bm k}$ Haldane model i.e., $H^{\uparrow}_{\bm k}=-(H^{\downarrow}_{-\bm k})^*$. As a result, the  perpendicular component of the spin, $\tau_z$ is conserved and the average magnetic field vanishes.  The real space representation of the Kane-Mele model is given by  
\begin{eqnarray}
\label{main:ham_Kane_Mele}
H &= & -t_1 \sum_{\langle ij \rangle} c^{\dagger}_i c_j + iV_R \sum_{\langle ij \rangle} c^{\dagger}_i (\vec{\tau} \times \hat{d}_{ij})_z c_j + M \sum_i c^{\dagger}_i \sigma_z c_i  \nonumber \\
&+& i V_{\rm so} \sum_{\langle\langle ij\rangle\rangle} \nu_{ij} c^{\dagger}_i \tau_z c_j
\end{eqnarray}
where $c_j(c^{\dagger}_j)$ represents four component fermionic annihilation (creation) operator consisting of spin and sub-lattice degrees of freedom. Here, $\vec{\tau}=(\tau_x,\tau_y,\tau_z)$ represents the spin degrees of freedom and $\vec{\sigma}=(\sigma_x,\sigma_y,\sigma_z)$ indicates the sub-lattice degrees of freedom. The first term is the same as the first nearest neighbour hopping in graphene.  In the second term with Rashba SOC, $\hat{d}_{ij}= (\hat{\bm{\delta}}_1,\hat{\bm{\delta}}_2,\hat{\bm{\delta}}_3)$ denotes the nearest neighbour unit vectors when electron traverses between the adjacent sites $j$ and $i$. The third term denotes the IS breaking mass term just like Haldane model. 
The last term for intrinsic SOC contains the chirality factor $\nu_{ij}$ defined by $\nu_{ij}=(\hat{\bm{\delta}}_l^{ij}\times \hat{\bm{\delta}}_m^{ij})_z =\pm 1$. The $\pm$ sign  depends on
the orientation of the two nearest neighbor unit vectors $\hat{\bm {\delta}}_l$ and $\hat{\bm{\delta}}_m$ along the bonds $l$ and $m$ that the electron traverses in going from site $j$ to $i$ where $l\ne m=1,2,3$.  
In the momentum space, the model reduces to the following form considering the basis $(c^{A\uparrow}_{\bm k},c^{A\downarrow}_{\bm k},c^{B\uparrow}_{\bm k},c^{B\downarrow}_{\bm k})^T$ 
\begin{equation}
    H_{\bm k}= \sum_{i=1}^5 d_{i} \Gamma_i + \sum_{i<j=1}^5 d_{ij} \Gamma_{ij}
\end{equation}
where $d_1=t(1+ 2 \cos x \cos y), d_2= M, d_3=V_{\rm so} (1-\cos x \cos y), d_4=-\sqrt{3} V_{\rm so} \sin x \sin y$, $d_{12}=-2 t \cos x \sin y, d_{15}= V_{\rm so}(2 \sin 2x - 4 \sin x \cos y), d_{23}=-V_R \cos x \sin y, d_{24}=\sqrt{3} V_R \sin x \cos y$ and $\Gamma_1=\sigma_x \otimes \tau_0, \Gamma_2=\sigma_z \otimes \tau_0, \Gamma_3=\sigma_y \otimes \tau_x, \Gamma_4=\sigma_y \otimes \tau_y, \Gamma_5=\sigma_y \otimes \tau_z$, $\Gamma_{i,j}=[\Gamma_i,\Gamma_j]/2i$. Here $x=k_x/2$, and $y=\sqrt{3}k_y/2$.

A close inspection suggests that  Rashba SOC, caused by a perpendicular
electric field or interaction with a substrate,  can support the topological phase even though it breaks mirror symmetry and spin conservation along $z$-axis. However, the intrinsic SOC, preserving $s_z$, is inevitable to obtain the topological phase even in the absence of the Rashba SOC.  One can find that the topological quantum spin Hall insulator phase persists for $V_R< 2\sqrt{3} V_{\rm so}$. On the other hand, the sub-lattice symmetry breaking mass term $M$ is another key factor to achieve the QSHI phase. To be precise, the topological phase is present for $3\sqrt{3}V_{\rm so} >M$ when $V_R=0$. Similar to the Haldane model, the sign of the gap changes  between Dirac points ${\bm K}$ and ${\bm K}'$ due to the $\pm \sigma_z \tau_z$ form of the intrinsic SOC. The $ V_{\rm so}$ term can be thought of as the second nearest neighbour $t_2$ term of the Haldane model except for the spin degrees of freedom. Note that the time reversal symmetry protected  topological phase is characterized by $Z_2$ index representing the spin-polarized edge modes. To be precise, spin up channel propagates in clock-wise manner while its time reversal partner spin down channel propagates in counter-clockwise manner on the edges indicating the SOC mediated spin momentum locking phenomena. 

\section{FOTI phases in honeycomb lattice}
\label{sec:FOTI-model}
Having discussed the Haldane and Kane-Mele model  in the previous Sec. \ref{sec:CI and QSHI}, we now demonstrate the model Hamiltonian that we consider for our study. As already mentioned, this model is an admixture of Haldane model of Chern insulator and Kane-Mele model of QSHI with appropriate ingredients \cite{Saha21}. The Hamiltonian is given by,
\begin{eqnarray}
\label{main:ham1}
H = && -t_1 \sum_{\langle ij \rangle} c^{\dagger}_i c_j + iV_R \sum_{\langle ij \rangle} c^{\dagger}_i (\vec{\tau} \times \hat{d}_{ij})_z c_j + M \sum_i c^{\dagger}_i \sigma_z c_i  \nonumber \\
&&~ + t_2 \sum_{\langle\langle ij \rangle\rangle} e^{i \nu_{ij} \phi} c^{\dagger}_i c_{j} + \frac{i V_{\rm so}}{\sqrt{3}} \sum_{\langle\langle ij\rangle\rangle}e^{i  \nu_{ij} \phi} \nu_{ij} c^{\dagger}_i \tau_z c_j 
\end{eqnarray}
where $c_i$ and $c^{\dagger}_i$ represent the four-component fermion spinors on $i$-th site. The model contains nearest neighbour spin-insensitive hopping of strength $t_1$ as the first term that includes no phase factor. 
The model contains a nearest neighbour spin-dependent hopping represented by $V_R$, the Rashba interaction as the second term which is the same as the Kane-Mele model. The third and fourth terms denote the IS and TRS breaking on-site mass term and the second nearest neighbour hopping with magnetic flux term, respectively which are the same as that of the Haldane model.  
The last term represents the  intrinsic SOC coupled to a chiral magnetic flux field. Therefore,  
for the second nearest neighbour we also have spin-independent hopping characterized by $t_2$ and also spin-dependent hopping represented by   $V_{\rm so}$, the spin-orbit coupling term.  The second nearest neighbour hopping contains the phase factor $e^{i \nu_{ij} \phi}$\cite{haldane-1988}. Here  $\nu_{ij}= \pm 1 $ as already described in the previous Sec. \ref{sec:CI and QSHI} ~\cite{kane-2005-2nd} which implies that the two different spin sectors have  opposite sign. Physically this is equivalent to two copies of Haldane model where each copy experiences a spin-dependent next-nearest hopping. 


\indent
{Now it is instructive to describe the main  features  of the Hamiltonian as described  in Eq. (\ref{main:ham1})  and explain the physical motivations of each term and their role in governing the topology of the system. Let us first investigate how to recover the original Haldane model  from Eq. (\ref{main:ham1}). This is easily done by setting  $V_R=V_{\rm{so}}=0$. Here only important to remember is that model reduced to two copies of Haldane model one for spin up and another for spin down and total Hamiltonian can be written as $H= H_{\uparrow} \otimes \mathbbm{1}  + \mathbbm{1}\otimes H_{\downarrow} $ with $H_{\uparrow}=H_{\downarrow}$. This equivalence of  $H_{\uparrow}=H_{\downarrow}$ points out that the  spin indices are irrelevant and  topological phases in each spin sector follow the condition  $M= 3 \sqrt{3} t_2 \sin \phi$ concomitantly. Here we observe that there is no distinction between the two spin sectors and thus the topological phases are expected  to be identical.} \\
\indent

Now let us consider another useful limit of the model where the two spin sectors are yet decoupled dynamically. This is obtained with $V_R=0$ and $t_2=0$. In this case the Hamiltonian can be written as before $H= H_{\uparrow} \otimes \mathbbm{1}  + \mathbbm{1}\otimes H_{\downarrow} $ where $H_{\uparrow} \neq H_{\downarrow}$ but their parametric dependence  is found to be $H_{\uparrow}(\pm |V_{\rm so}|) = H_{\downarrow}(\mp |V_{\rm so}|)$. To understand the emerging topology in this case, it is useful to compare the present case with ($V_{\rm so} \ne 0,t_2=0)$  with respect to the former case with ($V_{\rm so} = 0,t_2 \ne 0)$. The present case can be alternatively thought of as a magnetic flux being reversed from $\phi=\pi/2$ to $\phi=-\phi/2$ when spin up is converted to spin down. It might be said that TRS is virtually restored because under TRS the spin-up sector is mapped to spin-down sector with the reversed sign of magnetic flux or magnetic field. However, kinematically topological nature of each spin-sector is intimately tied with the spin-less Haldane model with flux $\phi$ fixed at $\pi /2$.
One can think of making $V_{\rm so}=0$ but having the magnetic flux associated with second nearest neighbour $\phi$ to be $\pi/2$ for one spin species and $-\pi/2$ for the other spin.   \\

\indent
Now we consider the more general case when with $V_{\rm so},V_R \neq 0$. We  note that in this case the Hamiltonian can not be written as the direct sum of two spin-sectors as there exist spin-dependent hopping processes where a spin flip transition between  neighbouring site occurs for electrons. However one may note that the Rashba SOC induced hopping does not break any TRS. The terms which are responsible for breaking TRS are  second neighbour hopping process determined by $t_2$ and $V_{\rm so}$. \\

\indent
Having described the various limits of Hamiltonian (\ref{main:ham1}), we write it down in the momentum space. To do that we incorporate the generalized $\Gamma$ matrices as suitably defined below 
\begin{equation}
H({\bm k})= \sum_{i=0}^9 n_i (\bm k) ~ \Gamma_i 
\label{main:hamk}
\end{equation}
with $\Gamma_i=\sigma_{i} \otimes \tau_0$ for $i=1,2,3$, $\Gamma_{i+3}=\sigma_{i} \otimes \tau_1$ for $i=1,2$, $\Gamma_{i+5}=\sigma_{i} \otimes \tau_2$ for $i=1,2$, $\Gamma_{8}=\sigma_{3} \otimes \tau_3$, $\Gamma_{9}=\sigma_{0} \otimes \tau_3$ and $\Gamma_{0}=\sigma_{0} \otimes \tau_0$. We interchangeably use  $\tau_{x,y,z}$ and $\sigma_{x,y,z}$ as $\tau_{1,2,3}$ and $\sigma_{1,2,3}$, respectively.  Here ${\bm \sigma}$ represents the orbital degrees of freedom and ${\bm \tau}$ corresponds to  spin degrees of freedom. Here we have used the basis $(c_{A\uparrow}, c_{A\downarrow}, c_{B\uparrow}, c_{B\downarrow})^T$ to write the Hamiltonian \ref{main:hamk}. Various $n_i$'s are given  as follows.

\begin{eqnarray}
\label{eqn01}
n_0= 2 t_2 f({\bm k}) \cos \phi , n_1=-t_1 (1 + 2 h({\bm k})   ), \\
\label{eqn2}
n_2= -2 t_1  \sin \frac{\sqrt{3}k_y}{2} \cos \frac{k_x}{2}, \\
\label{eqn3}
n_3= M - 2 t_2 ~g({\bm k}) \sin \phi , \\
\label{eqn4}
n_4= \frac{V_R}{\sqrt{3}} \sin \frac{\sqrt{3}k_y}{2} \cos \frac{k_x}{2} ,\\
\label{eqn5}
n_5= \frac{V_R}{\sqrt{3}} (h({\bm k}) - 1) , \\
\label{eqn6}
n_6= - V_R \cos \frac{\sqrt{3}k_y}{2} \sin \frac{k_x}{2} , \\
\label{eqn7}
n_7= V_R \sin \frac{\sqrt{3}k_y}{2} \sin \frac{k_x}{2}, \\
\label{eqn8}
n_8= \frac{V_{\rm so}}{3} g({\bm k}) \cos \phi , \\
\label{eqn9}
n_9= \frac{V_{\rm so}}{3} f({\bm k}) \sin \phi,
\end{eqnarray}

The various functionals which are present in the above equations are given below.
\begin{eqnarray}
\label{eqf}
 f({\bm k})= 2 \cos \frac{\sqrt{3}k_y}{2} \cos \frac{k_x}{2} + \cos k_x, \\
\label{eqg}
 g({\bm k})= 2 \cos \frac{\sqrt{3}k_y}{2} \sin \frac{k_x}{2} - \sin k_x ,\\
\label{eqh}
h({\bm k})= \cos \frac{\sqrt{3}k_y}{2} \cos \frac{k_x}{2}.
\end{eqnarray}
One may explicitly check the  presence (or absence) of TRS in different limits of the parameter regime by calculating
whether the identity $ \mathcal T H({\bm k}) \mathcal T^{-1} \ne  H(-{\bm k} )$ holds (or not). Here $\mathcal T = (I \otimes \tau_2)i \mathcal K$, $\mathcal K$ being the complex conjugation.

To understand the physical significance of each term of the extended Haldane model we note that in the original Haldane model the topology is determined mainly by the strength of second neighbour hopping $t_2$, the sub-lattice dependent mass term $M$ and the flux density $\phi$. These
terms connect the same-sub-lattice and spins and hence appear as the diagonal element of the resulting $2\times 2$  block matrix resembling Haldane model.  We expect that a great deal of understanding can be made by examining the diagonal terms of the resulting $4\times 4$ matrix for the extended Haldane model as well.  To this end, we investigate $H_{11}$ and $H_{22}$ components of the momentum space Hamiltonian which are given below.\\
\begin{eqnarray}  
\label{binup}
H_{11}&&= F(\bm k)_{\uparrow} \sin \phi + G(\bm k)_{\uparrow} \cos \phi  + M \\
\label{bindn}
H_{22}&&= F(\bm k)_{\downarrow} \sin \phi + G(\bm k)_{\downarrow} \cos \phi  + M \\
\label{fkup}
F(\bm k)_{\uparrow}&&= \bigg(f (\bm k) \frac{V_{\rm so}}{3} - 2 t_2 g (\bm k) \bigg) \\
\label{fkdn}
F(\bm k)_{\downarrow}&&= \bigg(-f (\bm k) \frac{V_{\rm so}}{3} - 2 t_2 g (\bm k) \bigg)\\
\label{gkup}
G(\bm k)_{\uparrow}&&= \bigg( g (\bm k) \frac{V_{\rm so}}{3} - 2 f(\bm k) t_2\bigg) \cos \phi  \\
\label{gkdn}
G(\bm k)_{\downarrow}&&= \bigg( -g(\bm k) \frac{V_{\rm so}}{3} - 2 f(\bm k) t_2\bigg) \cos \phi
\end{eqnarray} 
We have chosen above $H_{11}$ and $H_{22}$ such that the mass  term appears in identical sign i.e., they connect the same sub-lattices but correspond to different spins. This leads to a useful comparison between the two spin-sectors. We notice that in the absence of $V_{\rm so}$, $H_{11}=H_{22}$ as evident from Eqs. (\ref{binup}) and (\ref{bindn}).  In this context, one can consider a semiclassical treatment where the charges of electrons do matter but the spin degrees of electrons are not taken into account explicitly in the equation of motion. Consequently, the  electrons irrespective of their spins are affected the same way due to Lorentz force.  Thus the band inversion condition which is central for first order topological phases applies in the same footing for both the spins at the two Dirac points. However, we observe that when the spin-orbit interaction is considered, the band inversion conditions are determined by $F(k)_{\uparrow(\downarrow)}$ and $G(k)_{\uparrow(\downarrow)}$.   
Two spin sectors are distinctly influenced due to the change in sign of $V_{\rm so}$ as evident from Eqs. (\ref{fkup}) to Eq. (\ref{gkdn}). This scenario describes presence of  a Zeeman splitting dependent on $k$ which is different for up and down spin components or modulating the effective second neighbour hopping $t_2$ differently for each spin sector.  Thus we see that there is a competition between energy scales coming from the  orbital and spin degrees of freedom.  This implies that the band inversion condition at the topological phase transition point does not occur at the same $k$-points and associated Dirac cones are no longer interdependent.  In this section, we consider $t_1=1.0$ and $t_2=0.5$ without loss of generality. \\

\indent
\subsection{\textcolor{black}{Emergent topological  phases}}

Within the ambit of the Spin-Chern number we then introduce the following notation of $C_{\uparrow} $ and $C_{\downarrow}$ as the equivalent Chern number associated with the spin up and spin down eigenvectors. The many-body ground state projection operator is constructed from the negative energy eigenstates $|\phi_n(\bm k)\rangle$ of the Hamiltonian Eq. (\ref{main:hamk}): $P(\bm k)=\sum_{E_n<0}|\phi_n(\bm k)\rangle \langle \phi_n (\bm k)|$ where $E_n$ represents the $n$-th eigenvalue of $H(\bm k)$ in Eq. (\ref{main:hamk}). 
The spin projected  effective Hamiltonian is written in the above basis as follows $\tilde {H}(\bm k)= P^{-1}(\bm k) A P(\bm k)$ where  spin operator  $A=\sigma_0 \otimes \tau_3$. After diagonalizing $\tilde {H}(\bm k)$, we obtain four eigenstates $|\tilde{\phi}_n(\bm k)\rangle$ with energies $\tilde{\epsilon}_n(\bm k)$ where $\tilde{\epsilon}_1(\bm k)=-\tilde{\epsilon}_4(\bm k)<0$ and 
$|\tilde{\epsilon}_2(\bm k)| = |\tilde {\epsilon}_3(\bm k)| \simeq 0 $. One can define a spin spectrum gap $\Delta(\bm k)=\tilde{\epsilon}_4(\bm k)-\tilde{\epsilon}_1(\bm k)$ from the spin projected effective Hamiltonian \cite{spin-chern-number}. 
The finite nature of the spin gap $\Delta(\bm k) \ne 0 ~~\forall ~~{\bm k}$, allows us to construct the spin up Chern number $C_{\uparrow}$ and spin down Chern number $C_{\downarrow}$, following Fukui's formalism \cite{spin-chern2}, from four component eigenstates $|\tilde{\phi}_1(\bm k)\rangle$ and $|\tilde{\phi}_4(\bm k)\rangle$, respectively of $\tilde {H}(\bm k)$. Thus for a given set of parameters these two quantized invariant $(C_{\uparrow}, C_{\downarrow})$ will be used to define the topological phases.  The various terminology of phases has been discussed in the introductions can now be associated with this pair of Chern number and we define phases as below.

The QASHI phase is defined as when one of the spin-Chern number vanishes. Thus we have in total four  QASHI phase given by $(C_{\uparrow}, C_{\downarrow})= (0, \pm 1), (\pm 1,0)$. Next there is a possibility that the two spin-Chern numbers are opposite to each other and there are two such possibilities and we call these phases QSHI phases and characterized by    
$(C_{\uparrow}, C_{\downarrow})= (1, -1), (- 1,1)$. Finally, there is another possibility where the two spin-Chern numbers are finite yet equal to each other and there are two such possibilities given by  $(C_{\uparrow}, C_{\downarrow})= (1, 1), (- 1, -1)$. These phases are termed as QAHI phase. The reason for different nomenclature is such that whenever explicitly "spin" is mentioned such as in QASHI and QSHI we note that the edge current has explicit spin polarization. In the case of QASHI the edge current is charged and spin-polarized though in case of QSHI phase the edge current is charge neutral and spin-polarized. \textcolor{black}{However we would like to note that the usage of the word `anomalous', in QASHI might be confusing to some readers. This is because of the notion that regular and `anomalous' topology take place in the presence and absence of external magnetic field i.e., with broken and preserved TRS, respectively. 
Here we use the term `anomalous' in the spirit of taking the QSHI as the original phase which happens to be present in the absence of magnetic field. This  new phase with spin-current in the presence of magnetic field can also be termed anomalous QSHI as only one spin-component is topological unlike the regular QSHI. Nonetheless, this is a QSHI phase where only one spin degree of freedom is topological in the absence of the TRS. Therefore, technically QASHI phase is a TRS broken QSHI phase or a QSHE under TRS breaking}. 
On the other hand, QAHI phases have spin neutral charged current.
The above feature can be easily understood by the sign of the spin-Chern number yielding  different spin current directions. Thus whenever the signs of the spin-Chern numbers are identical for opposite spin-polarizations channels, they channel cancel each other yielding only spin-neutral charged current. Similar explanations can be found easily for QSHI and QAHI phases. For the future reference, we use $\rm{C}n$ and $\overline{\rm{C}}n$ with $n=1,2,3,4$ to represent these eight different phases with the following conventions of values of $(\rm{C}_{\uparrow}, \rm{C}_{\downarrow})$ as follows.  $\rm{C}1=(1,0)$, $\rm{C}2=(0,1)$, $\rm{C}3=(1,1)$, $\rm{C}4=(1,-1)$ and   $\overline{\rm{C}}1=(-1,0)$, $\overline{\rm{C}}2=(0,-1)$, $\overline{\rm{C}}3=(-1,-1)$, $\overline{\rm{C}}4=(-1,1)$ such that $\overline{\rm{C}}n=(-\rm{C}_{\uparrow}, -\rm{C}_{\downarrow})$ where  corresponding ${\rm{C}}n$ is defined as $(\rm{C}_{\uparrow}, \rm{C}_{\downarrow})$. Now we proceed to describe different phases that we just explained within the parameter spaces of extended Haldane model.


%
\begin{figure}[!htb]

 \includegraphics[width=0.47\linewidth]{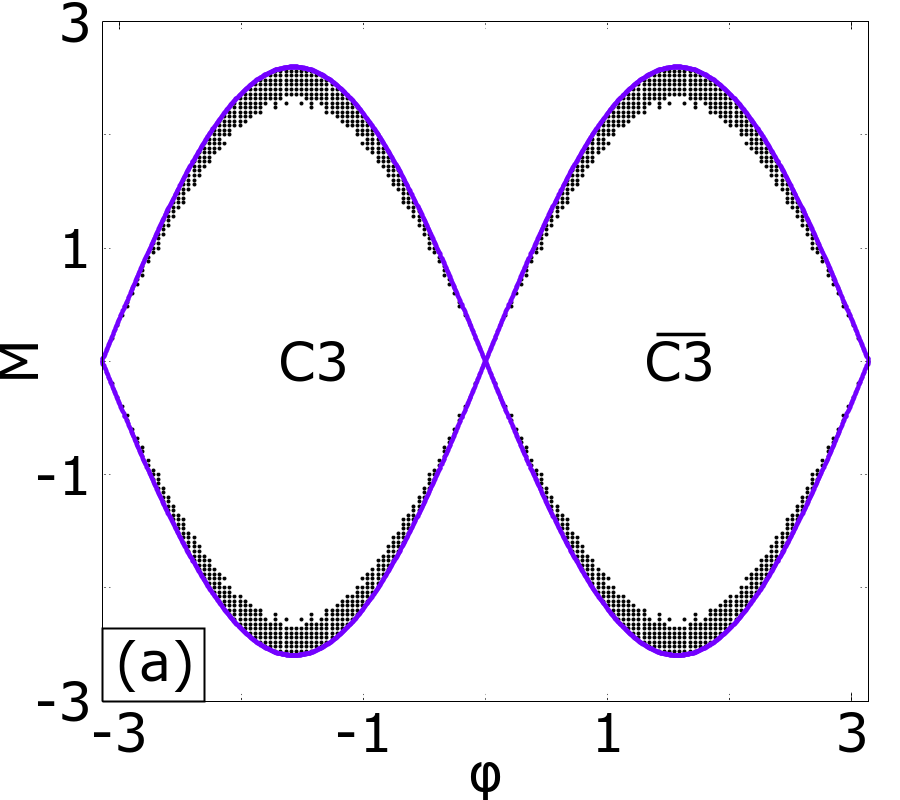}
 \includegraphics[width=0.47\linewidth]{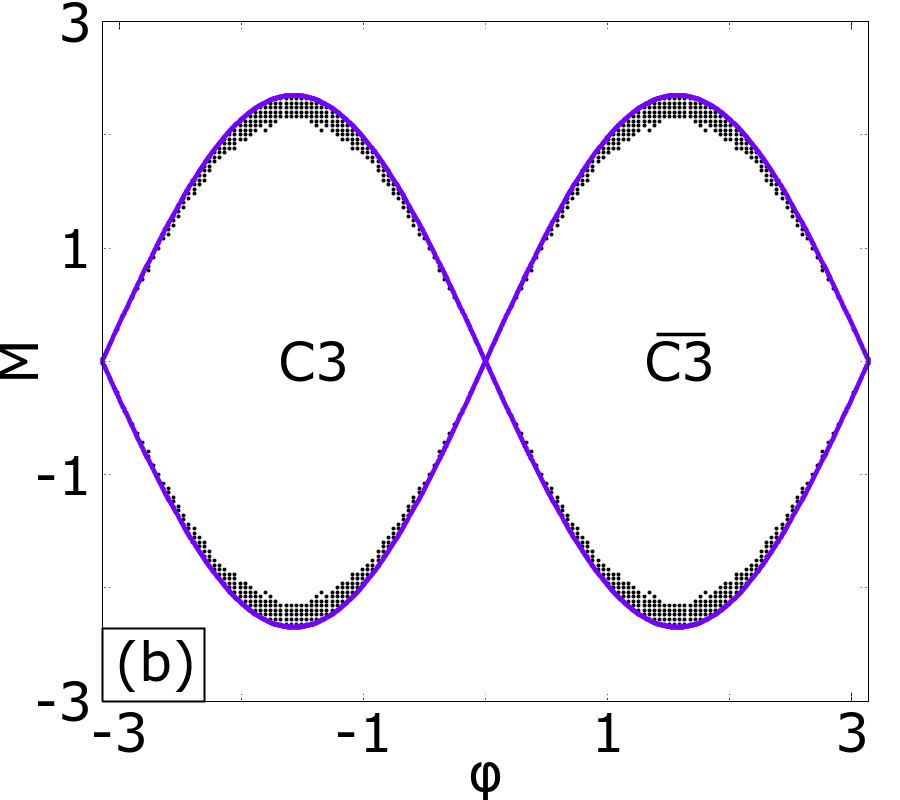}

\caption{Here we have drawn the phase diagram in $M-\phi$ plane for $V_R=1.0$ for two different values of $V_{\rm so}$. In (a) $V_{\rm so}=0.0$ and in (b) $V_{\rm so}=0.5$. The region which denoted by collection of black dots  is where spin-gap is vanishingly small. We note that as we increase the value of $V_{\rm so}$, this region gets thinner. After \cite{Saha21}.
}
\label{fig:modificationHaldane2}
\end{figure}



\subsection{\textcolor{black}{Roles of  $V_{\rm so}$ and $V_R$ in QAHI phase }}

We first discuss the simplest extension of Haldane model in spin-full system where the spin-Chern number is identically repeated in both the spin sectors which is characterized by $\rm{C}3$ and $\overline{\rm{C}}3$. We note that having $V_{\rm so}=0=V_R$ would automatically guarantee these phases as  the system just boils down to two identical copies of Haldane model. However, what is interesting is that a finite $V_R$ does not change the scenario which can be thought of as renormalization of hopping parameters on equal footing for both the spin sectors.  There is a finite extended range of values of $V_{\rm so}$ for which QAHI phase survives  as shown in 
Fig. \ref{fig:modificationHaldane2}. The phase boundary is plotted where the spin spectrum gap vanishes for a certain momentum mode. In Fig. \ref{fig:modificationHaldane2}, panel (a) and (b) we have taken $V_{\rm so}=0.0$ and $V_{\rm so}=0.5$ respectively while $V_R$ is taken to be unity.

\begin{figure}[!htb]

\includegraphics[width=0.23\textwidth]{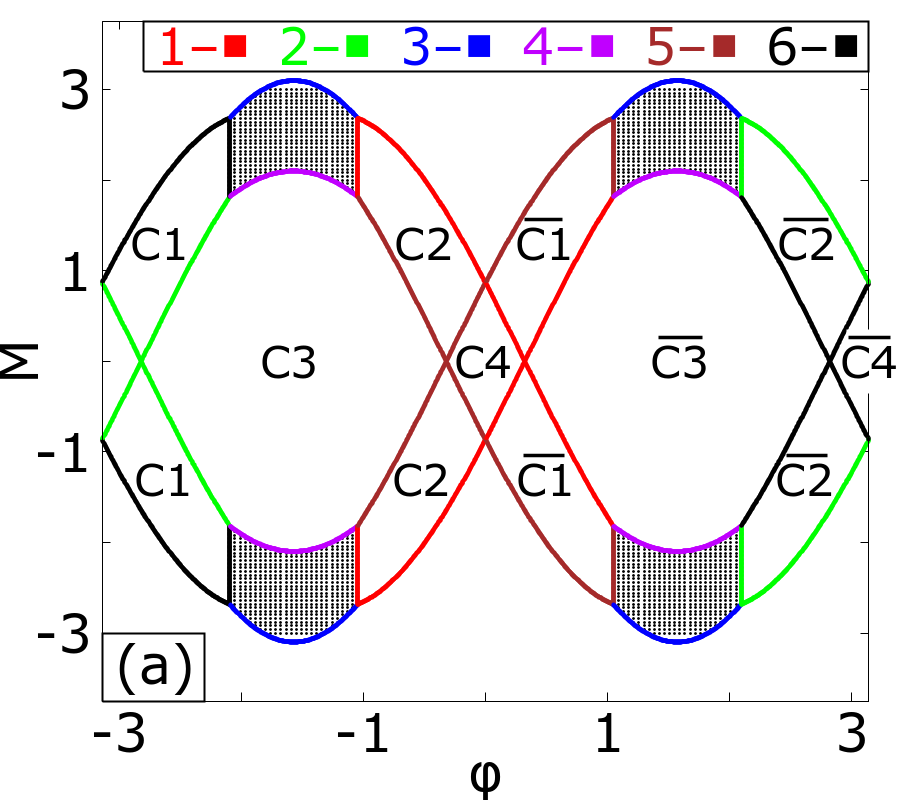}
\includegraphics[width=0.23\textwidth]{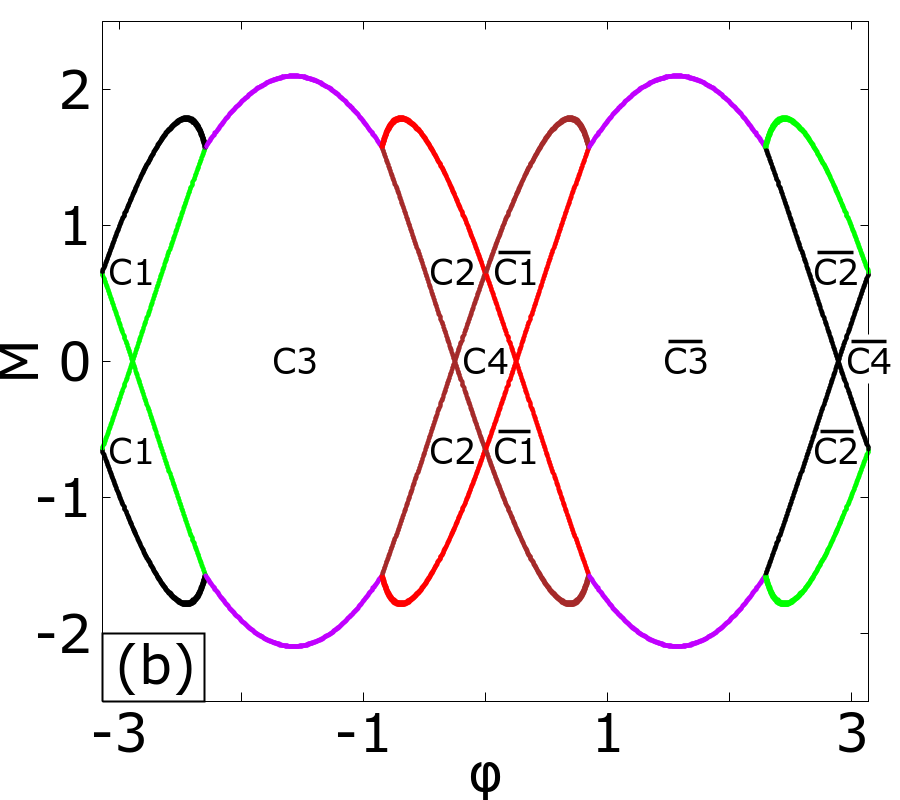}

\caption{In the above $V_{\rm so}$ is kept at 1.0 while for (a) and (b) $V_R=0.0$ and $0.5$ are taken respectively. 
The zeros of the band gap $\delta E_{n}^\alpha=0$ designate the following  phase boundaries: $ \delta E_{1}^\alpha = -\alpha M + \eta_{1,-} - \Delta_1 $ (red), 
$\delta E_{2}^\alpha = -\alpha M + \eta_{1,-} + \Delta_1$  (green), 
$\delta E_{3}^\alpha = 2\Delta_1$ (blue), 
 $\delta E_{4}^\alpha  = -2 \alpha M + \eta_{0,-}$ (magenta), 
$ \delta E_{5}^\alpha = -\alpha M + \eta_{-1,-} + \Delta_1 $ (dark brown),
$ \delta E_{6}^{\alpha} = -\alpha M + \eta_{-1,-} - \Delta_1 $ (black) with $\eta_{a,\pm}=\sqrt{3} aV_{\rm so} \cos \phi + \frac{1}{2}(V_{\rm so} \pm 6 \sqrt{3} t_2) \sin \phi$,   $  \Delta_1 =  \left( M^2+ 3V^2_R + (27t^2_2+ 3 \sqrt{3} t_2 V_{\rm so} + \frac{V^2_{\rm so}}{4}) \sin^2 \phi - 2 \alpha M \eta_{0,+} \right)$ where $\alpha=\pm$ denoted the Dirac points. The above gap equations are obtained  from four eigen-energies of the underlying low-energy model. As before the grey region denotes  a critical phase  marked by spin-gap close to zero, practically.  After \cite{Saha21}.}
\label{fig:modificationHaldane}
\end{figure}

The extended critical phase, represented by the collection of black dots, is a strikingly new finding, see Fig.~\ref{fig:modificationHaldane}(a). The spin gap vanishes identically for a momentum mode within this phase   \cite{spin-gap}. The spin-Chern number is unable to characterize this 
cap-like critical phase. For a fixed $V_R$ such that $V_R \leq V_{\rm so}$ the vertical width of the critical phase decreases as one increases $V_R$. On the other hand the horizontal width increases as $V_R$ is enhanced.  However when $V_R$ becomes larger than $V_{\rm so}$, then QASHI phases no longer exist with a concomitant  expansion of critical phases between  $-\pi<\phi< \pi$(the complete region bounded by the violet lines become critical for $V_{\rm so}=0$) as evident from the two panels of Fig. \ref{fig:modificationHaldane2}. For finite, as shown in Fig. \ref{fig:modificationHaldane2} (b) $V_{\rm so}$ QAHI phases decreases in size.  One may note that for $V_{\rm so} \geq V_R$ QASHI starts to appear around $\phi=0$. Remarkably, as one increases $V_R$ the phase boundaries (denoted by violet lines) expand. In this case,  the size of the critical phases increases and fully occupies the region in between $\pm \pi$. When $V_{\rm so}$ becomes larger than $V_R$ the critical phase is bounded by inside.  One can easily obtain the relation between $V_{\rm so}$ and $V_R$ by analysing the equations for vanishing gap at the phase boundary.



\subsection{\textcolor{black}{Roles  of $V_{\rm so}$ and $V_R$ in QSHI phases}}

Now we describe how the QSHI phase appears within the model Hamiltonian given in Eq. (\ref{main:ham1}). QSHI phase was first mentioned in the celebrated Kane-Mele model where the chirality of the edge modes of different spin-sectors are different. As a result, no net spin current is observed and hence the name suggests. The opposite chirality of edge states is governed the opposite Chern number of the two spin sectors. In the extended Haldane model the QSHI phase appears naturally for an extended region as marked by $\rm{C}4$ and $\rm \overline{C}4$ in Fig.~\ref{fig:modificationHaldane} where $V_{\rm so}$ is kept at 1.0 while for (a) and (b) $V_R=0.0$ and $0.5$ are taken respectively.  It is interesting to note that the QSHI phase appears for a narrow region width of small $M$ around zero and also for a narrow width around $\phi=0$. From Eqs. (\ref{binup}) and (\ref{bindn}), we find that $\sin \phi$ has small amplitude while $\cos \phi$ has a large value whose sign changes from $\phi=0$ to  $\pi$ explaining the origin of $\rm {C}4$ and $\overline{\rm C}4$  near $\phi=0$ and $\pi$, respectively. The opposite nature of spin-Chern number can be understood from Eqs. (\ref{gkup}) and (\ref{gkdn}) where $V_{\rm so}$ appears with opposite signs forcing band inversion conditions for two different spin sectors to be at different Dirac points.  From Fig.  Fig.~\ref{fig:modificationHaldane} (b), we notice that a finite $V_R$ shrinks the QSHI region a little and increases the  phase regions for QAHI denoted by ${\rm C}3$ and $\overline{\rm C}3$.

\subsection{\textcolor{black}{Complex interplay of parameters in QASHI phases}}

Finally we are in a position to the most intriguing topological phase where the spin-Chern number of one spin-sector is finite and the spin-Chern number of other spin sector is zero. This is marked by region $\rm{C}1$, $\rm{C}2$, $\overline{\rm{C}}1$, $\overline{\rm{C}}2$ in Fig. ~\ref{fig:modificationHaldane}. We notice that  various QSAHI phases appear at relatively large magnitudes of $M$ and near $\phi= \pm \pi/2$ and $\pi=\pm 3\pi/2$. From  Eqs. (\ref{binup}) and  (\ref{bindn}), we anticipate that the terms associated with $\sin \phi$ play a dominant role. This points to the fact that in the limit when  $V_R$ and $V_{\rm so}$ are both zero, the original Haldane model satisfies the condition for topological phase as $M/t_2 < 3 \sqrt{3} \sin \phi$. Now this condition implies that a comparatively large value of $M$ does not favor the topological phases. What we notice from 
Eqs. (\ref{binup}) and (\ref{bindn}) that $V_{\rm so}$ changes sign for two spin sectors indicating a dynamical increase and decrease in the effective mass of two spin sectors and hence favoring or destroying the topological phase. As a result, we obtain one spin sector appearing to be topological.
We believe that the Rashba SOC results in a detrimental effect in regard to the stability of the topological phases. The  shrinking of the topological region is also observed in Kane-Mele model if one increases $V_{R}$. Therefore, the effect of the Rashba SOC remains the same irrespective of the specific details of the model.

\subsection{\textcolor{black}{Topological phase diagram in $V_R - M$ plane}}

\begin{figure}[!htb]
 \includegraphics[width=0.48\linewidth]{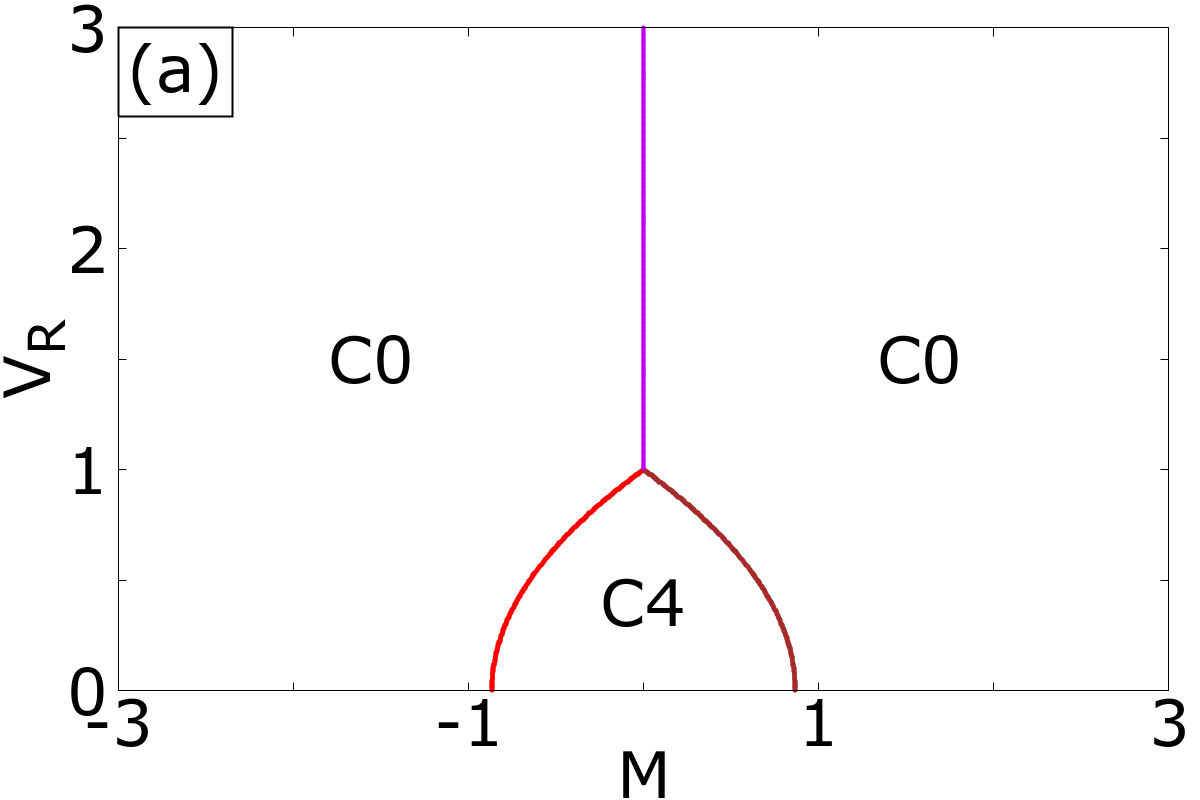}
 \includegraphics[width=0.48\linewidth]{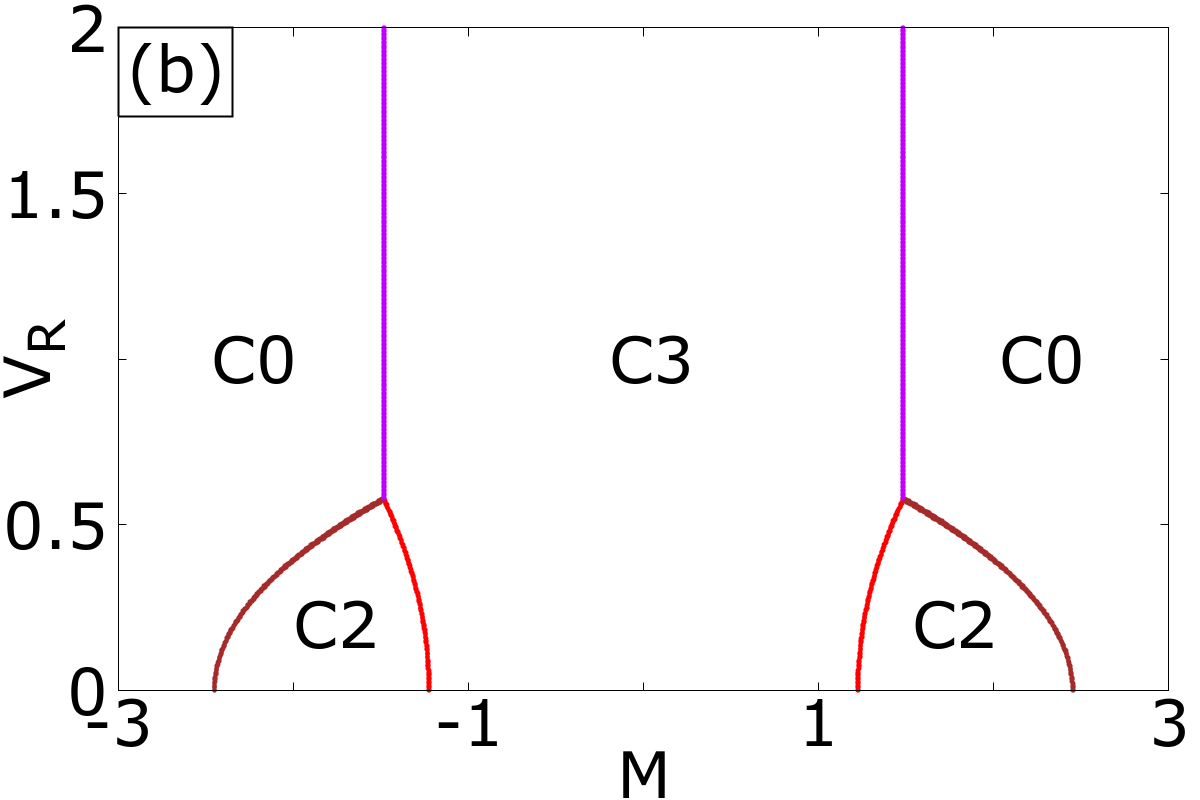}
\caption{ In the above  we show QAHI for $\phi=0$ in (a) and QASHI phase for $\phi=-\pi/ 4$ (b). For both cases $V_{\rm so}=1.0$ is taken. As evident vertical and horizontal axes represent $V_R$ and $M$ respectively. After \cite{Saha21}.
}
\label{fig:modificationKaneMele}
\end{figure}

In the Haldane model \cite{Haldane-88}, the role of  nearest neighbour hopping $t_1$ is only to provide a pair of Dirac points whereas topology is determined by second nearest neighbour hopping $t_2$, IS symmetry breaking mass term $M $ and the magnetic flux $\phi$. On the other hand, in Kane-Mele model \cite{kane-2005-1st,kane-2005-2nd}, the second nearest neighbour intrinsic SOC $V_{\rm so}$ and IS symmetry breaking mass term $M $ are mainly responsible for FOTI phases.  We also explain that the nearest neighbour spin-dependent hopping denoted by $V_R$ has  an important role in describing the topology for the spinfull case of Kane-Mele model. To clarify this we demonstrate the phase diagram in $V_R$-$M$ plane  exhibiting the changes over Kane-Mele phases.  The Figs.~\ref{fig:modificationKaneMele} (a) and (b) shows two QSHI phases for two values of $\phi$. In (a)  $\phi=0$ and  in (b) $\phi=-\pi/4$ without loss of generality. As evident for $V_R=0, \phi=0$, the model is reduced to two copies of Haldane model with modified second nearest neighbour hopping is evident from Eqs. (\ref{binup}), (\ref{bindn}),  (\ref{gkup}) and  (\ref{gkdn}). However, finite $V_R$ provides an effective second order hopping for each spin-sector (after integrating the other spin sector) which can alter the QSHI. This is the reason we observe that an increase of $V_R$  eventually destroyes the QSHI phase to a topologically trivial phase with a critical $V_R$ as seen in Fig. Fig.~\ref{fig:modificationKaneMele} (a).  Now we find that if we turn on a finite $\phi$, QSHI phase becomes unstable as shown in Fig.~\ref{fig:modificationKaneMele} (b), with $\phi=\pi/4$, QSHI is completely vanished and gives rise to two ${\rm C}2$s QASHI phase and one ${\rm C}3$ AQHI phase. In fact, the presence of a large region with QAHI \cite{sheng-2011} phase shows the most non-trivial effect of nearest neighbour Rashbha spin-orbit interaction denoted by $V_R$.

The phase diagram as shown in Fig. \ref{fig:modificationKaneMele} leads us to important insights.  While Fig. \ref{fig:modificationKaneMele}, (a) shows that QAHI phase makes transition to a trivial phase where both the spin-sector is topologically trivial. However, for QSHI phase, we find the appearances of QASHI phases for intermediate values of $M$.  This is also accompanied by the existence of multicritical points on the phase diagram where multiple topological phases touch each other. It is interesting that one can  define
$\Delta \rm{C}_{\sigma}$ for a given spin sector  $\sigma=\uparrow$, $\downarrow$, being the difference in $\rm{C}_{\sigma}$  among any  two  topological phases across a phase boundary.
It is found that $|\sum \Delta \rm{C}_{\sigma} |$ can only become unity over a phase boundary, demarcating two topological phases characterized by two sets of  $(\rm{C}_{\uparrow},\rm{C}_{\downarrow})$.  However  at multicritical point,  the above situation no longer generically holds true. The most important finding from the phase diagram can be considered the emergence of QASHI phase having only single  spin component out of two being topologically protected. One may note that for QSHI system obtained via doping magnetically is shown to lead to QASHI phases \cite{Liu-2008,li2013,qiao10}.  However,   QASHI found here  is the first example coming from a  tight binding model. The model Eq. (\ref{main:ham1}) remarkably hosts these phases naturally in the phase diagram.  \\
\indent
\noindent

\subsection{\textcolor{black}{Bulk boundary correspondence between topological edge mode and bulk invariant}}

\begin{figure}[!htb]

 \includegraphics[width=.95\linewidth]{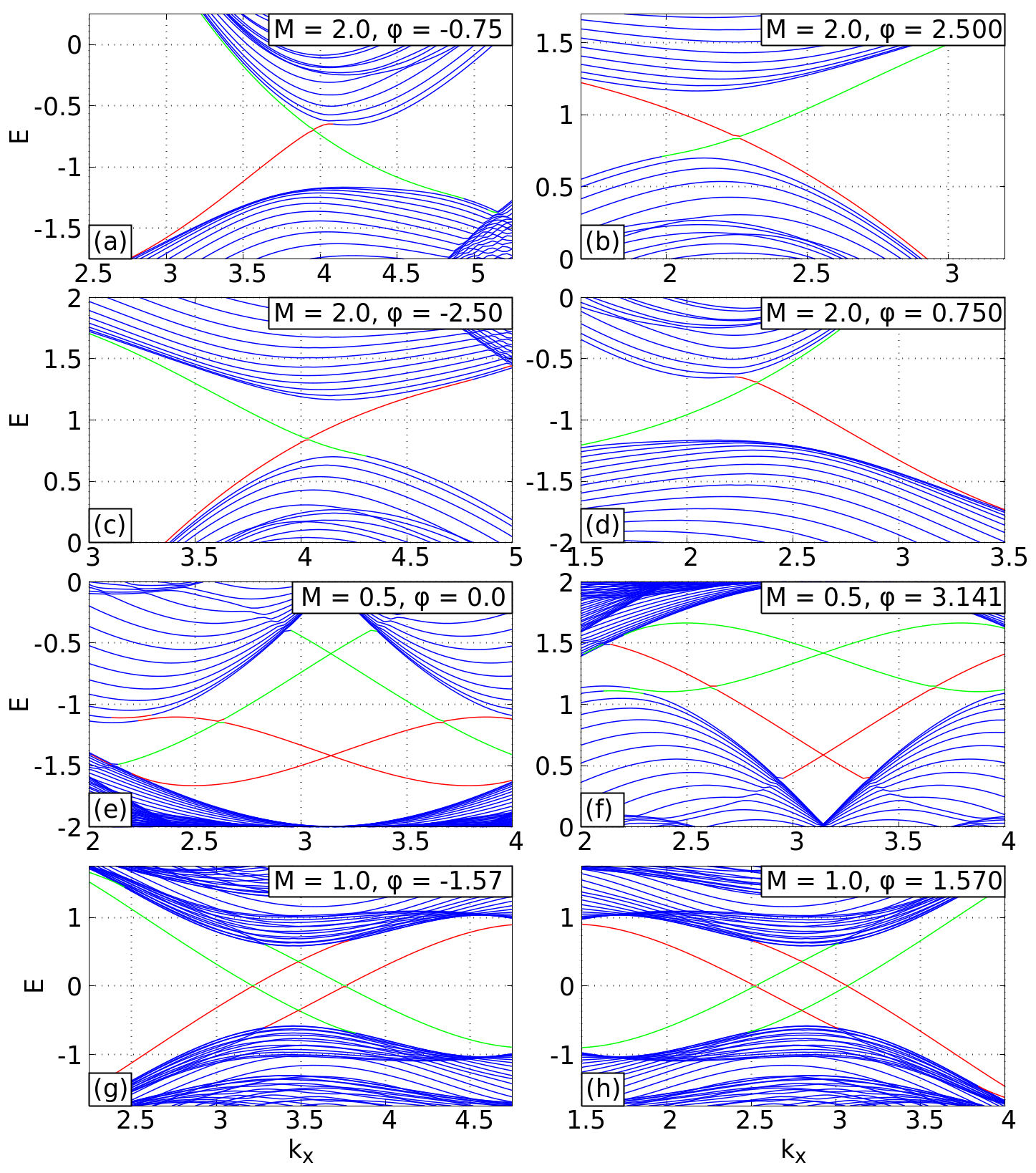}
 
\caption{ In the first four panel i.e., (a,b,c,d) the ribbon-geometry, adopting zig-zag semi-infinite chain, band structure of four QASHI are shown. This corresponds to one edge mode of a given chirality at one edge. (a) represents $(\rm{C}_{\uparrow},\rm{C}_{\downarrow})=(1,0)$, (b) for (-1,0), (c) for (0,1) and (d) for (0,-1).
The panel (e) and (f) correspond two QSHI phases. Panel (e) is with $(\rm{C}_{\uparrow},\rm{C}_{\downarrow})=(1,-1)$ while panel (f) is with $(-1,1)$. In a similar fashion (g) and (h) depict two QASHI phases such that $(\rm{C}_{\uparrow},\rm{C}_{\downarrow})=(1,1)$ for (g) and $(-1,-1$ for panel (h). 
After \cite{Saha21}.}
\label{fig:edgeStateVsoGTVrfoti}
\end{figure}

We are now in a position to discuss the nature of topologically protected edge states for various topological phases within this extended Haldane model. Note that this topologically protected edge states are obtained in a suitably defined open boundary system and for Honeycomb lattice it is the ribbon geometry with zig-zag edge we  consider\cite{nag19,Nag21,Ghosh21,Ghosh21b}. These special edge modes are often associated with certain chirality in the sense that they propagate along one direction on the edge. For ribbon-geometry, if one defines $N_{L}$ and $N_{R}$ as the left and right moving edge states, then the Chern number $C$ is defined as $\rm C=N_L-N_R $ \cite{haldane-1988, kane-2005-2nd}. For QASHI phase where the spin-Chern number is finite (and of magnitude unity) for only one spin sector, we thus obtain one edge mode per the open edges with opposite chirality. In Fig. \ref{fig:edgeStateVsoGTVrfoti} (a), (b), (c) and (d), we have plotted the various QAHI phases which depict the edge modes for $(1,0)$, $(-1,0)$, $(0,1)$ and $(0,-1)$ respectively. \\

\indent

In  Fig.~\ref{fig:edgeStateVsoGTVrfoti}(e) QSHI with $\rm(C_{\uparrow},C_{\downarrow})=(1,-1)$ has been considered to show band-structure for ribbon-geometry. While   Fig.~\ref{fig:edgeStateVsoGTVrfoti}(f) corresponds to the same with QSHI with $\rm(C_{\uparrow},C_{\downarrow})=(-1,1)$. Similarly Figs.~\ref{fig:edgeStateVsoGTVrfoti}(g)[(h)] corresponds to QAHI phase with  $\rm(C_{\uparrow},C_{\downarrow})=(1,1)[(-1,-1)]$. For all of these phases  both the spin sectors are topologically non-trivial and we obtain two edge modes at a given edge with appropriate chirality. For the two QSHI phases mentioned above with $\rm(C_{\uparrow},C_{\downarrow})=(1,-1)$ and $(-1,1)$ are such that the the chirality of the  two spin sector are opposite in both of these two phases. This results in charge neutral but spin-polarized current in the edge.  On the other hand for the two QAHI phases with $\rm(C_{\uparrow},C_{\downarrow})=(1,1)$ and $(-1,-1)$, the chirality of the both the spins are identical in a given phases which results in spin-neutral charge current.


\indent
One can appropriately construct a generalized bulk boundary correspondence for all these phases namely, QAHI, QSHI and QASHI phases by an extension of the definition \cite{haldane-1988, kane-2005-2nd},  $\rm{C}_{\sigma}= N^{\sigma}_{\rm RM}- N^{\sigma}_{\rm LM}$, where  $N^{\sigma}_{\rm RM}$ ($N^{\sigma}_{\rm LM}$) corresponds to the number of right (left) moving edge mode for spin $\sigma=\uparrow,\downarrow$ in a given edge. However due to the presence of $V_{\rm so}$, the numerical calculations do not show complete spin polarization for these edge modes. This makes the implementation of the above difference namely $\rm{C}_{\sigma}= N^{\sigma}_{\rm RM}- N^{\sigma}_{\rm LM}$ to be used here. To overcome this we define the edge states to be effectively up-spin polarized if its projection to up-spin axis is more than the down-spin projection. This effective description of spin-polarization then correctly captures the bulk-boundary correspondence. 


\subsection{\textcolor{black}{Low-energy analysis using the kinematics of the Dirac points}}
After having described the lattice model in detail, we now delve into exploring the low energy model. This would serve us to understand in deeper
perspective. To have study the low energy model one expands the model Hamiltonian  at the Dirac points $\alpha=\pm 1$. This yields, 
\begin{eqnarray}
\label{lwen1}
n_1&&= - \alpha \sqrt{3} t_1 k_x/2,\\
\label{lwen2}
n_2&&=\sqrt{3} t_1 k_y/2 , \\ 
\label{lwen3}
n_3&&= M + \alpha 3 \sqrt{3} t_2 \sin \phi \theta({\bf k}) , \\
\label{lwen4}
n_4&&=-V_R k_y/4 , \\
\label{lwen5}
n_5&&=- V_R/\sqrt{3} +\alpha V_R k_x/4 , \\
\label{lwen6}
n_6&&=\alpha V_R/\sqrt{3} + V_R k_x/4 ,\\
\label{lwen7}
n_7&&= -\alpha 3 V_R k_y/4 , \\
\label{lwen8}
n_8&&=-\alpha (\sqrt{3}/2) V_{\rm so} \cos \phi \theta(\bf k) ,\\
\label{lwen9}
n_9&&= - (V_{\rm so}/2) \sin \phi \theta({\bf k}).
\end{eqnarray}
where  $\theta({\bf k})= (1- |k|^2/4)$. The eigen-energies at the Dirac points has the form iven below,
\begin{eqnarray}
\label{lw14}
E_{1,4}&&= (w_1 + w_4 \pm  \lambda^{14}_1)/2, \\
\label{lw23}
E_{2,3}&&= (w_2 + w_3 \pm \lambda^{23}_2)/2
\end{eqnarray}
 with 
\begin{eqnarray}
\label{lww1}
w_1&&=n_3 + n_8 + n_9 , \\
\label{lww2}
w_2&&=n_3-n_8-n_9,\\ 
\label{lww3}
w_3&&=-n_3 -n_8 + n_9\\
\label{lww4}
w_4&&=-n_3 +n_8 - n_9,\\
\label{lwr1}
r_1&&=-n_5 - n_6, \\  
\label{lwr2}
r_2&&=n_6-n_5, \\
\label{lwlambda}
\lambda^{jk}_i &&=\sqrt{4 r^2_i + (w_j-w_k)^2}. 
\end{eqnarray}
We below analyze the above mathematical expressions for certain cases. 

\begin{center}
{\bf $V_R=0$ case}
\end{center}


As mentioned earlier the band inversion criteria have to be met for any spin sector to become topological and it is instructive to first discuss the  simple case of $V_R=0$. For $V_R=0$, the   energy gap for spin up is given by $\Delta E^{\uparrow}_{AB}=w_1-w_3$, and for spin down  $\Delta E^{\downarrow}_{AB}= w_2 -w_4$ which can be obtained from Eqs. (\ref{lw14}) and (\ref{lw23}) (where A refers conduction band and B refers valence band). One may note that the Bernevig-Hughes-Zhang model for HgTe Quantum Well \cite{bhz1} is closely related to this low energy model. This implies that one may proceed to examine different phases in an equivalent way. The criteria of band inversion implies that for a given   topological phase i$\Delta E^{\sigma}_{AB}$ have opposite signs at the two Dirac point ${\bm K}$ and ${\bm K}'$.  Thus we arrive at the governing condition
as follows,
\begin{eqnarray}
\label{bandin}
&&\Delta E^{\sigma}_{AB} ({\bm K}) \Delta E^{\sigma}_{AB} ({\bm K}') <0
\end{eqnarray}
where $\sigma= \uparrow {\rm or} \downarrow$.  Different topological phases are obtained for different way of satisfying the above criteria. For the QASHI phases where only one  spin sector is topological the above criteria is satisfied for one spin-sector and the remaining spin sector is trivially gapped out such that $\Delta E^{\sigma}_{AB}$  has same sign at both the Dirac points.   On the other hand the QSHI and QAHI  phases satisfy the above criteria for both the spin sectors. The criteria given in  Eq. (\ref{bandin}) also clearly demonstrates that there are two possible QSHI (and QAHI) phases. The main difference between QSHI and QAHI phases is that in QSHI phases band inversion condition of both the spin sectors are opposite at two Dirac points whereas it is identical for QAHI phase at the two Dirac points.\\

\indent
We note from Eqs. (\ref{lwen5}) and  (\ref{lwen6}) that $n_5=n_6=0$ for $V_R=0$. In that case one can further simplify the $\Delta^{\sigma}_{AB}{\bm{k}_i}= M + \zeta 3 \sqrt{3} t_2 \sin \phi + \xi (V_{\rm so}/2) \sin \phi$ using Eqs. (\ref{lw14}), (\ref{lw23}),  (\ref{lwen3}), and  (\ref{lwen9}) with expressions of $w_i$ as given from Eqs. (\ref{lww1}) to (\ref{lww4}). Here $\zeta, \xi$ can take $\pm$ depending on which Dirac point and which spin sector we are interested in. Thus one can define $x_{\zeta, \xi}$ so that,
  \begin{equation}
  x_{\zeta,\xi}=M + \zeta 3 \sqrt{3} t_2 \sin \phi + \xi (V_{\rm so}/2) \sin \phi.
  \end{equation} 
The above quantity would lead us to obtain the phase boundary between various topological phases. When the spin up channel is topologically non-trivial then we have $x_{+-}x_{-+}<0$. Similarly when the spin-down sector is topological  i.e  $\rm{C}_{\downarrow} \ne 0$, we have  $x_{++}x_{--} < 0$. To obtain the phase boundary separating regions with different values of $\rm{C}_{\uparrow}$ , one solves   $x_{\pm\mp} $= 0 for M. For the down-spin sectors the corresponding equation is $(x_{\pm\pm})$=0. It also clarifies why    $\rm{C}_{\sigma}$   can only changes by unity at phase transition. However we note that  $\rm{C}_{\sigma}$  can change by more than unity  at the multicritical points where at least three or more phases converge including non-topological phases. For $V_R =0$, one can find various topological phases  as illustrated in  Fig.~\ref{fig:modificationHaldane} and Fig.~\ref{fig:modificationKaneMele} by the low energy analysis  described  above.


\begin{center}
{\bf $V_R \ne 0$ case}
\end{center}

We now examine low energy analysis   for finite $V_R \ne 0$. It is revealed that  $V_R, V_{\rm so} \ne 0$ and $\phi \ne 0$, only adjust the phase boundary to some extent. As a result all the topological phases do survive as before, only the  area of a topological phases for which it occurs decrease or increase. Thus we can infer that all the topological phases that exist without $V_R$ can be transported  adiabatically as $V_R$ is turned on. This imply that the symmetry constrant governing the topological phases do not changes.  Representing $y_{\eta}= \eta \sqrt{3} (V_{\rm so}/2) \cos \phi$ and $z_{\zeta,\xi}=\sqrt{ 4 V_R^2/3 + x^2_{\zeta,\xi}}$, obtained from Eqs. (\ref{lwen8}), (\ref{lw14}), (\ref{lw23}) and (\ref{lwlambda}), we find \\
\begin{eqnarray} 
\Delta E^{\uparrow \downarrow}_{AB}({\bm k}_1)&& = \pm  y_{-} + |x_{+,-}| {\rm sgn}(x_{+,-}) + z_{+,+} \\
\Delta E^{\uparrow \downarrow}_{AB}({\bm k}_2)&& = \pm y_{+} + |x_{-,+}| {\rm sgn}(x_{-,+}) + z_{-,-}.
\end{eqnarray}
This allows us to write the band inversion condition of the spin up and down sectors as given below. 
\begin{eqnarray}
\label{bnup}
&&[y_{-} + |x_{+,-}| {\rm sgn}(x_{+,-}) + z_{+,+}] \nonumber \\
&&[y_{+} + |x_{-,+}| {\rm sgn}(x_{-,+}) + z_{-,-}] < 0~~~ {\rm and}\\
\label{bndn}
&& [y_{+} + |x_{+,-}| {\rm sgn}(x_{+,-}) + z_{+,+}] \nonumber \\&&[y_{-} + |x_{-,+} |{\rm sgn}(x_{-,+}) + z_{-,-}] < 0 
\end{eqnarray}
where Eq. (\ref{bnup}) refers spin up sector and Eq. (\ref{bndn}) refer to the spin down sector. In both cases, the condition $z_{\zeta,\xi}^2>0$ is necessary to be complied with to obtain the phase boundary.  Physically  it translates into modifying the phase boundaries for $V_R \ne 0$.  However, one should note that  the competition between  $V_{\rm so} \sin \phi$ and $\sqrt{3}V_R/2$ terms also plays important roles  in shaping the phase boundaries. To mention an example of this, we find that the phase boundary is markedly altered yielding  a  new topological phase  for $\phi \ne 0$ (which corresponds to TRS broken system) as depicted in Fig.~\ref{fig:modificationKaneMele} (a) and (b). 
One can define the spin-Chern number in an effective manner, considering the bulk band gap structure at two Dirac points,  as follows 
\begin{eqnarray}
 \rm{C}_{\sigma}=&&\frac{1}{2} \Bigg[{\rm sgn} \bigg(\Delta E^{\sigma}_{AB}({\bm k}_2)\bigg) - {\rm sgn} \bigg(\Delta E^{\sigma}_{AB}({\bm k}_1)\bigg) \Bigg]
 \label{sc_number}
\end{eqnarray}
 with
$\sigma=\uparrow,\downarrow$.  As for QAHI phases one needs to examine the spin gap  $\Delta \mathcal{E}^{\uparrow \downarrow}_{A}= E^{\uparrow}_A - E^{\downarrow}_A $, and $\Delta \mathcal{E}^{\uparrow \downarrow}_{B}= E^{\uparrow}_B - E^{\downarrow}_B $ at two Dirac points. The stability of topological phase depends on the robustness of the non-zero spin gap.  Which implies that we must have $\Delta \mathcal{E}^{\uparrow \downarrow}_{A}, \Delta \mathcal{E}^{\uparrow \downarrow}_{B} \ne 0$, \cite{spin-gap} to obtain finite  $(\rm{C}_{\uparrow},\rm{C}_{\downarrow})$. Interestingly, our numerics in lattice model indicates it is possible spin gap vanishes at any points within the  momentum BZ and need not be identical at which physical spectrum get vanished i.e at Dirac points.


\section{SOTI phases in honeycomb lattice}
\label{SOTI-model}


\begin{figure}[!htb]
\includegraphics[width=.95\linewidth]{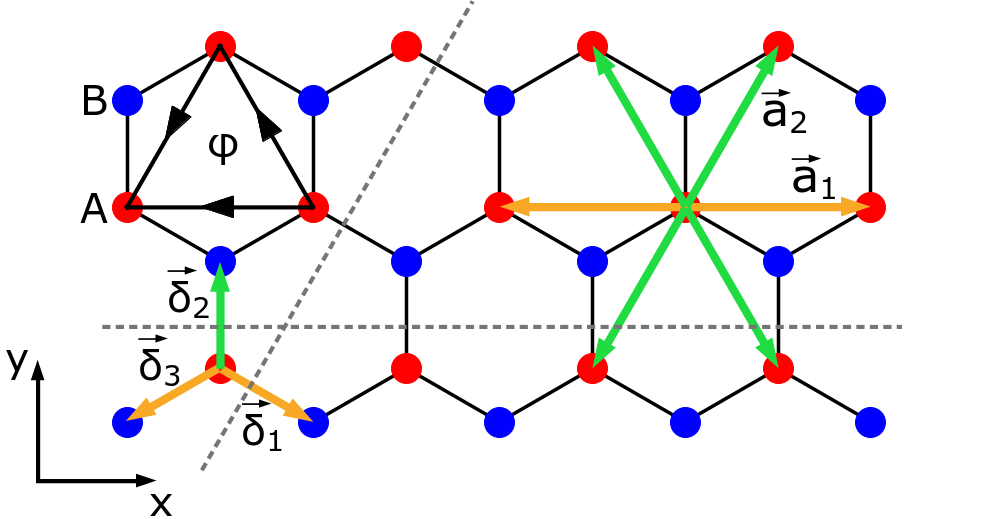} 
\caption{The graphene honeycomb lattice structure with nearest neighbour links ${\bm \delta}_{1,2,3}$ and primitive  lattice vectors ${\bm a}_{1,2}$. We show the triangular structure of the magnetic flux plaquette $\phi$ connecting the second nearest neighbour sub-lattices.  }
\label{fig:anisotropygraphene}
\end{figure}


Having discussed the FOTI phases in the previous Sec. \ref{sec:FOTI-model}, based on the modified Haldane and Kane-Mele model, we now demonstrate the SOTI phases of the above model. 
As discussed in Sec.\ref{intro} to realize the HOTI phases it is essential to reduce the discrete lattice rotational symmetries of the model. For example, in square lattice and triangular lattice the four-fold and six-fold rotational symmetries are reduced to two-fold and three-fold rotational symmetries \cite{schindler-2018,benalcazar-2017,benalcazar-2017-prb,parames-2017}, respectively,  by introducing anisotropy in hoopping strength. This anisotropy is reminiscent of a one-dimensional SSH model where the magnitude of hopping strengths of adjacent bonds are not equal rendering the bi-partite nature of the lattice. For hexagonal lattice, an anisotropic hopping scheme  is adopted in Haldane model \cite{Wang21}.

Let us first describe the anisotropy as depicted in Fig. \ref{fig:anisotropygraphene}. As shown, for the nearest neighbour hopping, the vertical hopping is made stronger in comparison to the bonds on zig-zag chains extended along horizontal direction. Similarly, for the  second nearest neighbour hopping, the slanted bonds are made stronger in comparison
to the second nearest neighbour hopping along horizontal direction. In both cases, the ration between the  magnitude of weaker and stronger bonds is denoted by $\eta$. $\eta=1$ denotes the isotropic limit and $\eta=0$ denotes a fully anisotropic limit. Note that even for $\eta=0$, the lattice does not disintegrate into one dimensional chain due to the non-vanishing third nearest neighbour hopping. The lattce formed in this case is also shown in Fig. \ref{fig:anisotropygraphene} right panel. The generalized Haldane model with broken  $C_3$ symmetry\cite{Wang21,Saha21,haldane-1988,kane-2005-1st,saha-23} is given below.
\begin{eqnarray}
\label{main:ham}
	H = && -\sum_{\langle ij \rangle} t^{ij}_1 c^{\dagger}_i c_j + \sum_{\langle\langle ij \rangle\rangle} t^{ij}_2 e^{i \nu_{ij} \phi} c^{\dagger}_i c_{j} + M \sum_i c^{\dagger}_i \sigma_z c_i  \nonumber \\
	&& + ~ \frac{i V_{\rm so}}{\sqrt{3}} \sum_{\langle\langle ij\rangle\rangle}e^{i \nu_{ij} \phi} \nu_{ij} c^{\dagger}_i \sigma_z c_j + g \sum_i c^{\dagger}_i \tau_z c_i
\end{eqnarray}
where $c_i(c^{\dagger}_i)$ indicates the  creation (annihilation) operator
with ${\bm \sigma} \in (A,B)$ and ${\bm \tau} \in (\uparrow,\downarrow)$ representing the orbital and spin degrees of freedoms.

One may notice that the Hamiltonian in Eq. (\ref{main:ham}) is similar to the Hamiltonian proposed in Eq. (\ref{main:ham1}) with the following differences. First, the nearest neighbour and second nearest neighbor hopping denoted by $t_1$ and $t_2$ are made anisotropic which will be explained shortly. Secondly for simplicity, we do not include  Rashba interactions denoted by $V_R$ term. However, it may  give additional HOTI physics that we leave for future studies. Finally in Eq. (\ref{main:ham}), we consider a Zeeman term which was absent in Eq. (\ref{main:ham1}). We note that in both the Eqs. (\ref{main:ham1}) and (\ref{main:ham}) SOC represented by $V_{\rm so}$ denote a spin-dependent second nearest neighboor hopping  and is made isotropic in both the cases. As the nearest neighbour hopping amplitide $t_1$ is now bond dependent it is written as $t_1^{ij}$ and will be represented by a vector $t^{ij}_1=(\eta t_{1}, t_{1},\eta t_{1})$ with $|\eta|<1$ where the first, second and third entries denote the hopping amplitude in ${\bm \delta}_1$, ${\bm \delta}_2$ and ${\bm \delta}_3$ directions. The expressions of $\delta_i$ are given as ${\bm \delta}_1=(1/2,-1/2\sqrt{3})$, ${\bm \delta}_2=(0,1/\sqrt{3})$ and ${\bm \delta}_3=(-1/2,-1/2\sqrt{3})$ as already discussed in Sec. \ref{sec:CI and QSHI}. Therefore, ${\bm \delta}_{1,3}$ link hosts weak bond while ${\bm \delta}_2$ represents strong bond.
The spin-independent second nearest neighbour anisotropic hoppings are represented by $t^{ij}_2=(\eta t_{2}, t_{2}, t_{2})$ along $ {\bm a}_1=(1,0)$, ${\bm a}_2=(1/2,\sqrt{3}/2)$ and ${\bm a}_3=(-1/2,\sqrt{3}/2)$ as shown in Fig. \ref{fig:anisotropygraphene}.
In order to obtain the SOTI phase, one can employ two types of cut.  Cut through the vertical bond ${\bm \delta}_2$ designate a weak bond cut while cut through the slanted bond ${\bm \delta}_1$ represents the strong bond cut. The weak (strong) bond cut is indicated by cut 1 (2) as depicted in the Fig. \ref{fig:anisotropygraphene} with a slanted (horizontal) dashed line.
The $C_3$ symmetry is broken due to 
the factor $|\eta| \ne 1$ (with $|\eta| < 1$). This is manifested in the strong and weak bonds of strengths $t_{1,2}$ and  $\eta t_{1,2}$, respectively. Here, $M$ represents the IS breaking sub-lattice mass term while  $g$ is responsible for the breaking of TRS due to the magnetic field acting on spin-degrees of freedom. The definition of  $\nu_{ij}$ is already described in Sec. \ref{sec:CI and QSHI} \cite{Saha21,kane-2005-1st}.

The momentum space Hamiltonian,  obtained from Eq.~(\ref{main:ham}), in the basis $(c_{A\uparrow}, c_{A\downarrow}, c_{B\uparrow}, c_{B\downarrow})^T$ is given by \cite{saha-23} 
\begin{eqnarray}
\tilde{H}({\bm k},\eta)= \sum_{i=0}^5 \tilde{n}_i  ~ \tilde{\Gamma_i} 
\end{eqnarray} 
with $\tilde{\Gamma}_i=\sigma_{i} \otimes \tau_0$ for $i=1,2,3$, $\tilde{\Gamma_{4}}=\sigma_{3} \otimes \tau_3$, $\tilde{\Gamma_{5}}=\sigma_{0} \otimes \tau_3$ and $\tilde{\Gamma_{0}}=\sigma_{0} \otimes \tau_0$.
Note that in comparison to the definition of $\Gamma$ matrices as defined after Eq. (\ref{main:ham1}), $\tilde{\Gamma}_4$ and $\tilde{\Gamma}_5$ are defined differently for the sake of convenience.  The components $\tilde{n}_i$ are given by:
\begin{eqnarray}  
&&\tilde{n}_0= 2 t_2 \tilde{f}({\bm k},\eta) \cos \phi , \\
&&\tilde{n}_1=-t_1 (1 + 2 \tilde{h}({\bm k},\eta)   ) , \\
&&\tilde{n}_2= -2 \eta t_1  \sin \frac{\sqrt{3}k_y}{2} \cos \frac{k_x}{2} ,\\
&&\tilde{n}_3= M - 2 t_2 ~\tilde{g}({\bm k},\eta) \sin \phi , \\
&&\tilde{n}_4= \frac{V_{\rm so}}{3} \tilde{g}({\bm k},1) \cos \phi , \\
&&\tilde{n}_5=g + \frac{V_{\rm so}}{3} \tilde{f}({\bm k},1) \sin \phi.
\end{eqnarray}
Here  $\tilde{f}({\bm k},\eta), \tilde{g}({\bm k},\eta)$ and $\tilde{h}({\bm k},\eta)$ are defined as follows.
\begin{eqnarray}
&&\tilde{f}({\bm k},\eta)= 2 \cos \frac{\sqrt{3}k_y}{2} \cos \frac{k_x}{2} + \eta\cos k_x, \\
&&\tilde{g}({\bm k},\eta)= 2 \cos \frac{\sqrt{3}k_y}{2} \sin \frac{k_x}{2} + \eta \sin k_x , \\
&&\tilde{h}({\bm k},\eta)= \eta \cos \frac{\sqrt{3}k_y}{2} \cos \frac{k_x}{2}. 
\end{eqnarray}


\begin{figure}[!htb]
\centering
\includegraphics[width=0.48\textwidth]{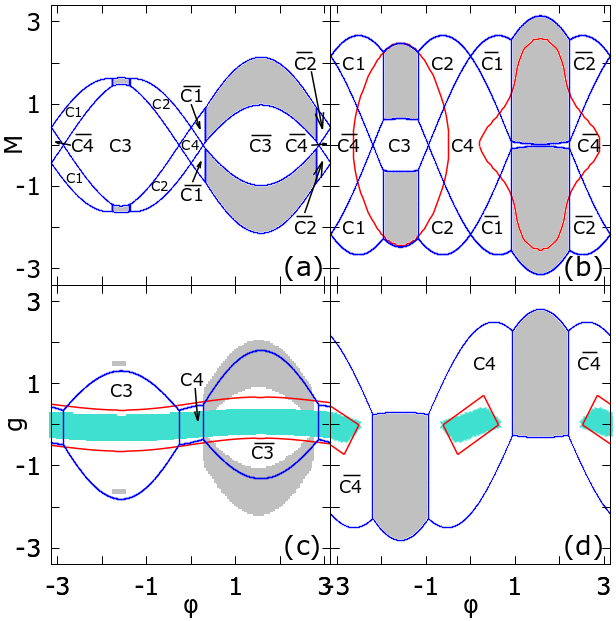}
\caption{We illustrate the phase diagram in $M-\phi$ plane in (a) and (b)  for $V_{\rm so} = 0.5$ and $2.5$, respectively with  $g=-0.33$.
In $g-\phi$ plane for $V_{\rm so} = 0.5$ and $2.5$, we demonstrate the phase diagrams in (c) and (d), 
respectively with  $g=-0.33, M=0$. We depict the spin spectrum gap (shaded gray region, red lines), bulk dipole moment (turquoise-colored region) and phase boundaries (blue lines).   After \cite{saha-23}.}	
\label{fig:1}
\end{figure}

As the starting Hamiltonian for investigating the  SOTI phases in the extended Haldane model as given in Eq. (\ref{main:ham}), we first discuss the fate of FOTI phases (in the limit $\eta=1$) in the presence of Zeeman term denoted by $g$.
In the case of 
TRS broken FOTI phases, spin-Chern number \cite{spin-chern2,prodan-2010} serves as a good topological index. This  is numerically computed with Fukui method \cite{Saha21}.
There exist a $\phi \rightarrow -\phi$ symmetry in the phase diagram in the absence of the Zeeman field $g=0$. To be precise, the  eight-fold FOTI phases can symmetrically appear under $\phi \rightarrow -\phi$ \cite{Saha21}. Interestingly, a finite Zeeman field $g\ne 0$ destroys the above symmetry by shrinking or expanding different phases.  \\
\indent

\begin{center}
{\bf \small {\textcolor{black}{ $g$-induced  modulation of QSHI and QAHI phase boundaries }}} 
\end{center}

Examining the  Figs.~\ref{fig:1} (a), (b) we note that the area of QSHI phases (denoted by $\rm{C}_4$ and $\bar{\rm C}_4$) for positive values of  $\phi$  is identical to the QSHI phase appearing for  negative values of $\phi$ \cite{saha-23}. This is also the case in the absence of Zeeman term as seen in Fig. \ref{fig:modificationHaldane2}. On the contrary, though the area of QAHI phase was identical for positive and negative regions of $\phi$  in the absence of Zeeman term, we note that in the presence of Zeeman term the QAHI phase (denoted by $\bar{\rm C}_3$ and $\rm{C}_3$ ) has a larger area for negative value of negative values of $\phi$ in comparison to positive values of $\phi$. \\
\indent

\begin{center}
{\bf \small {\textcolor{black}{$V_{\rm so}$-induced  modulation of QSHI and QAHI phase boundaries }}} 
\end{center}

The SOC denoted by $V_{\rm so}$, causes to increase the relative area for QSHI phases in comparison to QAHI phase as evident in  Fig.~\ref{fig:1} (b). It is interesting to note that a careful choice of $V_{\rm so}$ results in removing the QAHI phase for positive values of $\phi$. It is remarkable to note that the overall size of QASHI regions (denoted by $\bar{\rm C}_1, {\rm C}_1$ and $\bar{\rm C}_2, {\rm C}_2$) also increase as we increase the value of $V_{\rm so}$. Moreover, QSHI and QASHI phases remain symmetric under $\phi \rightarrow -\phi$. \\
\indent


\begin{center}
{\bf \small \textcolor{black}{ Comparison of  spin-spectrum gap between  SOTI and  FOTI phases}}
\end{center}

The spin-spectrum gap plays a decisive role in deciding the phase boundary across two FOTI phases. What we observed
while discussing various FOTI phases in the extended Haldane model is that there exists a region where spin-dpectrum gap is vanishingly small but yet not zero \cite{saha-23}. This region is non-topological as topological invariant can only be calculated in the presence of a well-developed gap. The vanishingly small spin-spectrum gap remains an interesting aspect of FOTI phases and here we discuss its possible role if any in determining the SOTI phases.  
We note that  the shaded gray region which denotes a vanishingly small spin-spectrum gap without any bulk topological index has a larger area  as we increase the value of $V_{\rm so}$, see Figs.~\ref{fig:1} (a), (b). 
The gray-shaded region denotes the finite value of the  spin-spectrum gap for $\eta=1$. This   region  is in complete accordance with the FOTI phase boundaries marked by blue lines. In Figs.~\ref{fig:1} (b), (c), (d) we superpose the spin-spectrum gap for $\eta=0.25$ on top of the FOTI phases obtained for $\eta=1$.
In order to illustrate the profile of the spin spectrum gap for $\eta=0.25$, 
we draw the red lines within which the 
spin spectrum gap remains finite. Interestingly, the region of finite spin-spectrum gap for $\eta=0.25$ and $\eta=1$ lie in significantly different portions of the phase diagram.  What is remarkable is that the SOTI phases have 
finite spin-spectrum gap  and are 
characterized by mid-gap corner states with the turquoise-colored region, see Figs. ~\ref{fig:1} (c) and (d)). 
One can thus infer that the SOTI phases are 
crucially dependent on the nature of the spin-spectrum gap. 
The SOTI phases are found to be embedded in QAHI and QSHI (QSHI) for $V_{\rm SO}=0.5$ ($V_{\rm SO}=2.5$). This is mentioned below when we describe Figs.~\ref{fig:1} (c) and (d). 



\begin{figure}[!htb]
\includegraphics[width=.95\linewidth]{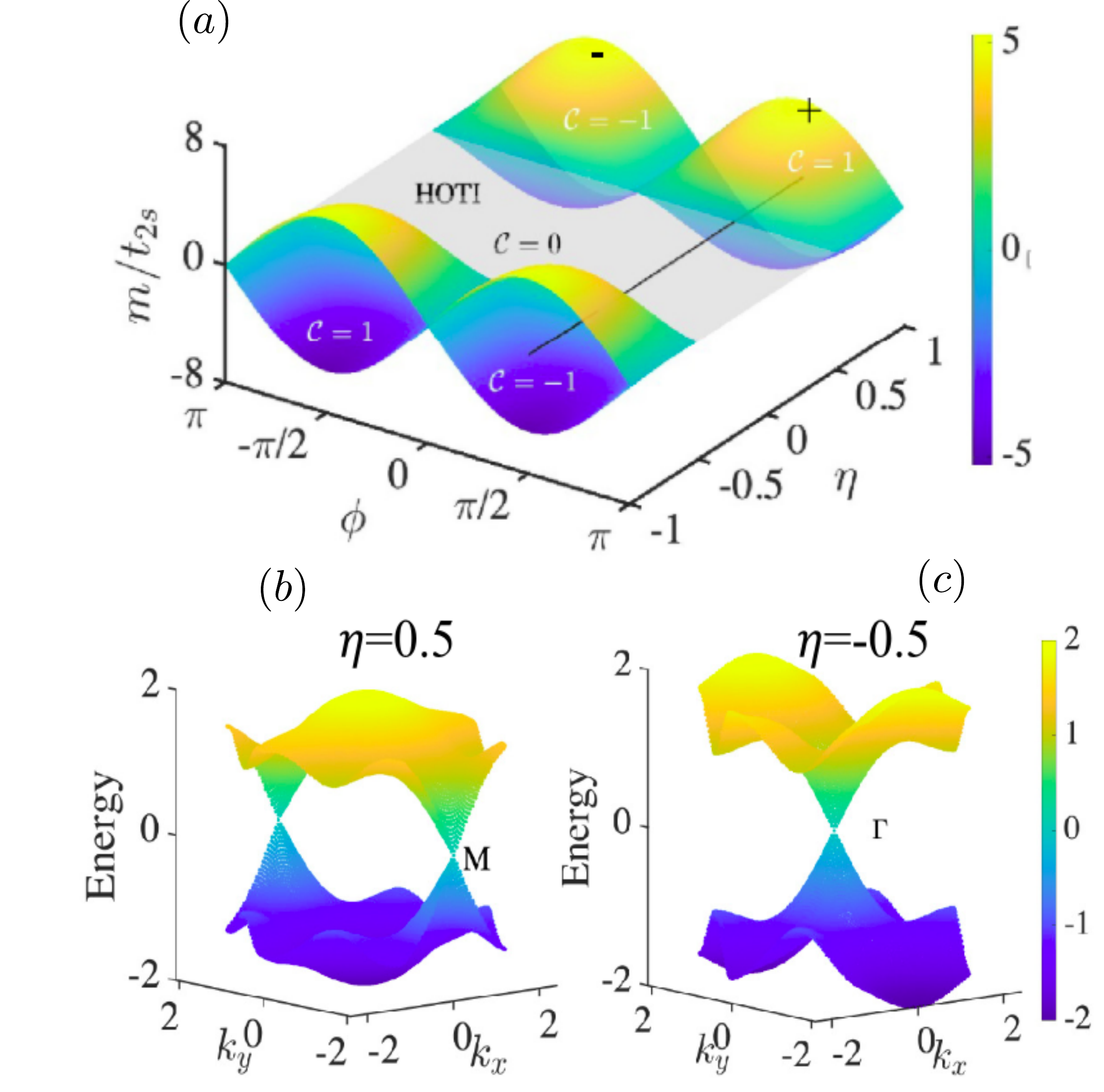}
\caption{Three parameter phase diagram in sub-lattice mass ($m$)-magnetic flux ($\phi$)-hopping anisotropy ($\eta$) plane  
is demonstrated in (a) showing the FOTI and HOTI phases for $0.5 \le |\eta|\le 1$ and $|\eta|<0.5$, respectively. The Dirac points in the BZ for $\eta=0.5$ and $-0.5$ are shown in (b) and (c), respectively. After \cite{Wang21}. }
\label{fig:banshil_dirac_node}
\end{figure}

 \begin{center}
{\bf \small \textcolor{black}{ Comparison of  gap closing condition between  SOTI and  FOTI phases}}
\end{center}

The topological phase transition from a topologically trivial phase to a non-trivial phase is accompanied by a gap closing at the critical point and reopening at either side of the critical points. If the trivial phase is a normal insulator, we would expect the gap to open in this region also. In the Haldane model, having $C_3$ symmetry and being an example of a FOTI, there exist two gapped Dirac points at ${\bm K}$ and ${\bm K}^{\prime}$ points except for the critical gapless lines. In the topological phase,  one Dirac point is positively i.e., trivially gapped while the other Dirac point is negatively i.e., topologically gapped. Once the model undergoes a phase transition to a trivial phase from the topological phase,  the band gap closes at the critical points and again re-opens at both the Dirac points that are now positively gapped.  This fact remains unchanged for the extended Haldane model given in Eqs. (\ref{main:ham1}) and (\ref{main:ham}) with $\eta=1$. The  Dirac points always stay at ${\bm K}$ and ${\bm K}^{\prime}$ is a direct consequence of $C_3$ symmetry of the model. It is now pertinent to discuss the nature of gap closing conditions and the appearances of Dirac points for such $C_3$ symmetry broken system. For this purpose we review the discussions made in the previous study \cite{Wang21} of $C_3$ symmetry broken Haldane model yielding SOTI phases. \\

\indent
The Fig. \ref{fig:banshil_dirac_node} describes how an anisotropic Haldane model affects the movements of Dirac points and associated gap closing and reopening at the transition from the FOTI phase to a SOTI phase. The limit $\eta=\pm 1$ corresponds to isotropic limit  where the Chern insulator phase exists depending on the flux $\phi$ and the ratio of sub-lattice dependent mass $M$ to the second nearest neighboor hopping $t_{2s}$. As $\eta$ is varied from $-1$ to $1$, one faces several distinct behavior. For our purpose, it is useful to divide the range of $\eta$ into three distinct regions. Region ${\rm I}$ extends between $-1\le \eta \le -0.5$ , region ${\rm II}$ for $-0.5 < \eta < 0.5$ and region ${\rm III}$ for $0.5 \le \eta \le  1$, see Fig.\ref{fig:banshil_dirac_node}, (a). Region ${\rm I}$ and ${\rm III}$ correspond to FOTI phase where two  Dirac points move away from ${\bm K}$ and ${\bm K}^{\prime}$ points due to broken $C_3$ symmetry. Upon changing $|\eta|$ from $1$ to $0.5$, in the region ${\rm I}$, the Dirac points move toward ${\bm M}$ and ${\bm M}^{\prime}$ points while in the region ${\rm III}$, both the Dirac points move toward ${\bm \Gamma}$ point as shown in Fig. \ref{fig:banshil_dirac_node} panel (b) and (c). For the region ${\rm II}$, the gap again opens up yielding a SOTI phase  with zero energy corner modes. We note that this SOTI phase only exists for $M=0$ i.e., IS preserved case. \\
\indent

Similar to the scenario described above the gap closing condition remains identical for the model Hamiltonian given in Eq. (\ref{main:ham}) and SOTI phase appears in the region ${\rm II}$ only. For region ${\rm I}$ and ${\rm III}$ i.e for $0.5<|\eta|<1$, two Dirac points appear at $\myvec{K}^{\eta}_{\pm} = ( \pm 2\theta,  -2\sqrt{3}\theta )$ with $\theta={\rm arctan}(\sqrt{4 \eta^2 -1})$.
The phase boundaries can be calculated when the gap vanishes for $H(\myvec{K}^{\eta}_{\pm},\eta)$. It is useful to write down the eigenvalues at these two Dirac points as given below

\begin{eqnarray} 
 &&\lambda^{\pm}_1=\tilde{n}^{\pm}_3-\tilde{n}^{\pm}_4-n^{\pm}_5, \\ 
&&\lambda^{\pm}_2=-\tilde{n}^{\pm}_3+ \tilde{n}^{\pm}_4-\tilde{n}^{\pm}_5, \\
&&\lambda^{\pm}_3=-\tilde{n}^{\pm}_3-\tilde{n}^{\pm}_4+\tilde{n}^{\pm}_5, \\
&&\lambda^{\pm}_4=\tilde{n}^{\pm}_3+\tilde{n}^{\pm}_4+\tilde{n}^{\pm}_5 
\end{eqnarray}
where various $\tilde{n}$'s are defined as follows.
\begin{eqnarray}
&&\tilde{n}^{+}_3= M + 6 t_2 \sin \phi \sin u, \\
&&\tilde{n}^{+}_4=  -(2V_{\rm so}/3) \cos \phi \sin u (1-\cos u), \\
&&\tilde{n}^{+}_5=  g+ (2V_{\rm so}/3) \sin \phi [(\cos u + 1/2)^2 -3/4], \\
&&\tilde{n}^{-}_3= M + 4 t_2 \sin \phi \sin u\\
&&\tilde{n}^{-}_4=  -(2V_{\rm so}/3) \cos \phi \sin u \\
&&\tilde{n}^{-}_5=  g- (V_{\rm so}/3) \sin \phi
\end{eqnarray}
In the above we use $\cos u= -1/2\eta$. Note that when $\eta=0.5$ ($-0.5$), two Dirac points reside at the ${\bm M}$ (${\bm \Gamma}$)-points \cite{Wang21}.  We are interested in the SOTI  for $0<\eta<0.5$, for which the Dirac points are gapped out and one obtains $\cos u =0,\sin u =1$. This yields  the effective bulk band-gap  $\Delta^{\pm}_{ij}=\lambda^{\pm}_i-\lambda^{\pm}_j$ can be of  various forms such as $(\Delta^{-}_{12},\Delta^{+}_{14})=(2M+8 t_2 \sin \phi -(4 V_{\rm so}/3) \cos \phi,-2 g+(2 V_{\rm so}/3) \sin \phi-(4 V_{\rm so}/3) \cos \phi)$.
This further suggests the emergence of multiple FOTI phases as observed in Fig. \ref{fig:1}.  The anisotropy by turning $\eta$ away from unity can non-trivially modify the phase boundary.  Hence it is 
pertinent to investigate the emergence of SOTI phase out of the
underlying  FOTI phase or a trivial phase. Interestingly,  similar to the $C_3$ symmetry broken Haldane model \cite{Wang21}, here also we find that SOTI phase originates from FOTI phase, namely from the QSHI phase as shown in Figs.~\ref{fig:1} (c) and (d).

\indent
One uses bulk-dipole moment $p_{\alpha}=-(1/2\pi)^2\int d^2 {\bm k} {\rm Tr}[A_{\alpha}]$ ($\alpha=x,y$) with non-Abelian Berry connection $[A_{\alpha}]^{mn}=-i\braket{u_{\bm k}^m | \partial_{k_\alpha} | u_{\bm k}^n}$ to topologically characterize the 
SOTI phase. Here the integration is over the entire BZ and $|u_{\bm k}^n \rangle$ represents $n$-th occupied band  associated with the Hamiltonian \cite{Ezawa18,Ezawa18b,benalcazar-2017}. For one-dimensional crystals,  adopting the concept of Wannier centers and Wilson loop, the  polarization can take the form 
$p_{\alpha}(k_\beta) =  -i\log \det \left[ \mathcal{W}_{\alpha, {\bm k}} \right]/2\pi$ with the Wilson loop ${\mathcal W}_{\alpha,{\bm k} }=F_{ {\bm k}+ (N_\alpha -1) \Delta k_{\alpha} } \cdots F_{ {\bm k} + \Delta k_{\alpha} } F_{\bm k} $. Here,  $ \left[F_{\bm k}\right]^{mn}=\langle u_{\bm k + \Delta k_\alpha} | u_{\bm k} \rangle$, and $\Delta k_\alpha= 2\pi /N_\alpha$ ($N_\alpha$ represents the number of discrete points considered inside the BZ along $k_\alpha$). Extending this concept to a two-dimensional crystal,    the total polarization along $\alpha$, terms as bulk-dipole moment,  can be found as $p_\alpha = \sum_{k_\beta} p_\alpha(k_\beta)/N_\beta$. The corner modes are characterized by $p_y=0.5$  and $p_x \ne 0.5$ (modulo unity) \cite{Wang21,saha-23}. For example, 
in the case of the IS-preserved modified Haldane model without the sub-lattice mass term $M=0$, the SOTI phase is characterized by a half-integer quantized dipole moment. 
This continues to hold for 
the present case with spin-full modified Haldane model in the presence of IS 
where the SOTI phases show the half-integer quantization.  
For finite $M\ne 0$, representing the IS broken case, the half-integer
quantization of $p_y$ is expected to be violated. This is what we examine when we investigate the SOTI phases for various parameter regimes. For all the remaining figures except  Fig.~\ref{fig:1},  we consider $\eta=0.25$  unless otherwise specified.

Now we are in a position to discuss the bulk-boundary correspondence for the SOTI phase i.e.,  
the appearances of corner modes and its connection with the bulk-dipole moment. For this purpose, it is useful to refer to Figs.~\ref{fig:1} (c) and (d), respectively, plotted for  $V_{\rm SO} = 0.5$, and $2.5$ with $M=0,\eta=0.25$, where the SOTI phase is designated by turquoise-colored region which is accompanied with a finite spin-spectrum gap as well as mid-gap corner states. The connection between the 
half-integer quantization and spin-spectrum gap at $\eta=0.25$ becomes evident when the quantization of $p_y$ is observed for all $\phi$ values  except at $\phi=0,\pi$. 
Within the above regions, the spin-spectrum gap is finite except at $\phi=0,\pi$. This is remarkable because for $\phi=0, \pi$, the FOTI phase becomes trivial and it reinforces the fact that FOTI phase is needed for the emergence of SOTI phase upon the onset of anisotropy. Importantly, for small $V_{\rm SO}=0.5$, we show  in Fig.~\ref{fig:1} (c) that SOTI phase, characterized by $p_y=0.5$,  emerges out of the underlying  QAHI and QSHI phases. For $\eta=0.25$, the vanishingly small spin-spectrum gap washes out a part of the FOTI phase leading to the  SOTI phase boundary to be flanked as compared to the FOTI phase boundary.

On the other hand, the QAHI phase vanishes  in the phase diagram
for relatively large $V_{\rm SO}=2.5$.
This causes QSHI phase only to give rise to the SOTI phase as depicted in Fig.~\ref{fig:1} (d). However, the SOTI phases
submerge in different FOTI phases. These SOTI phases happen to be characteristically same as far as their spatial distribution of the mid-gap states and bulk-dipole moment are concerned.
The corners of red line coincide with the boundary associated with turquoise-colored region in Fig.~\ref{fig:1} (d) 
for $g=0$ case only. For finite $g\ne 0$,
the mismatch between the red line and turquoise-colored region 
can be understood by an  apparent breakdown of generalized bulk-boundary correspondence in the $C_3$ symmetry-broken SOTI system. This  deserves further examination for its actual reason to be examined. The  model, considered here, breaks chiral symmetry and hence there does not exist any condition for the energy of the corner states. We now proceed to discuss how the SOTI phase manifests its different characteristics on specific parameters.



\subsection{\textcolor{black}{$M=0$ case with inversion symmetry}}

\begin{figure}[!htb]
\includegraphics[width=0.48\textwidth]{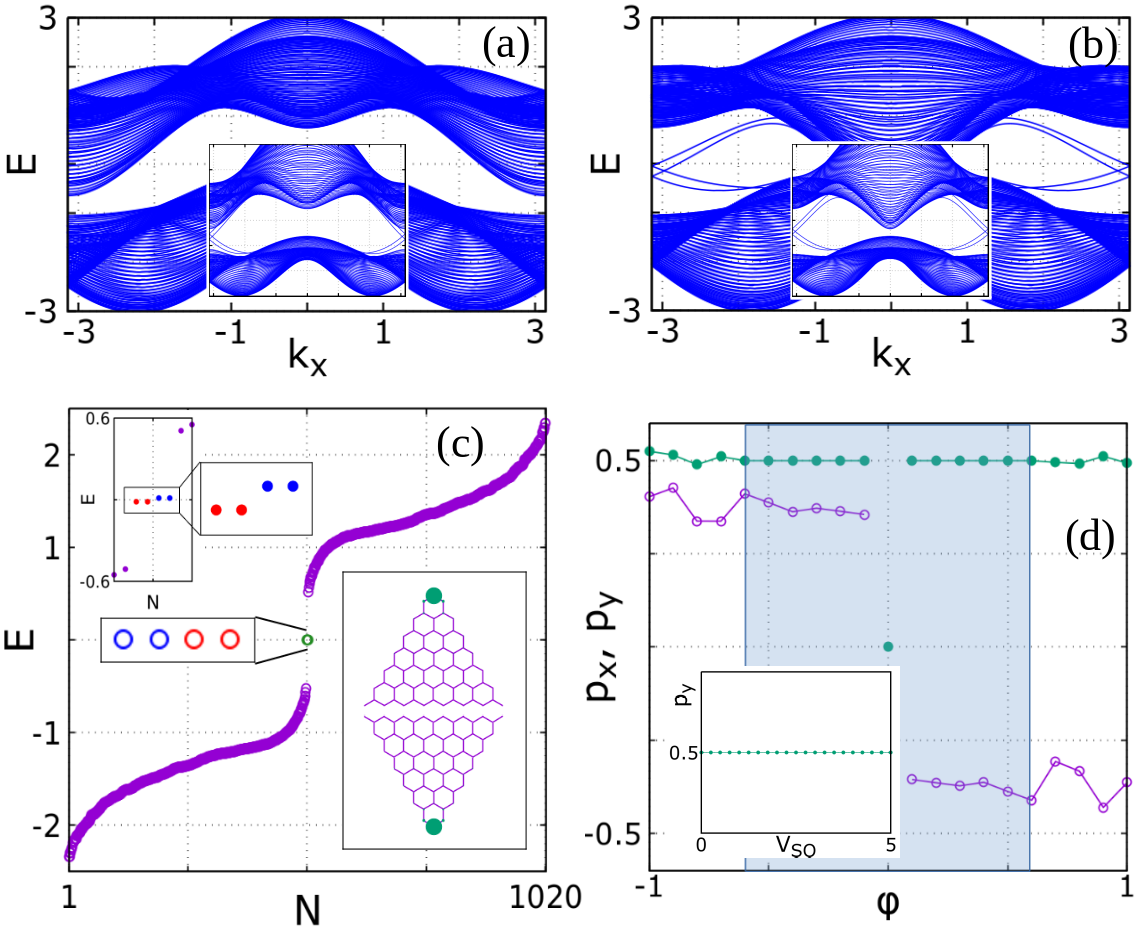}
\caption{Figures (a) and (b) depict the  ribbon geometry dispersion along cut 2 and cut 1, respectively, with  $\eta=0.25$. 
We repeat the above dispersion for $\eta=0.75$
in insets  (a) and (b). The band energy profile for cut 2 nano-disc geometry is demonstrated in (c).  Lower inset (c) displays the corner localization of mid-gap states in nano-disc for $g=-0.023$; upper inset (c) shows the energy of mid-gap states for $g=-0.01$. In  (d) the green solid-circle (pink empty-circle) denotes the evolution of $p_y$($p_x$) for $g=-0.023$, inset displays the change with $V_{\rm so}$. Parameters used are as follows: $M=0$, $\phi = 0.1$, and $V_{\rm so} = 2.5$. After \cite{saha-23}.}	
\label{fig:2}
\end{figure}

We begin with the scenario $M = 0, V_{\rm so} \neq 0, g \neq 0$ to investigate the SOTI phase. For this parameter set, the corner localization is shown by 
the mid-gap modes representing the  SOTI phase as depicted in Fig.~\ref{fig:1} (c) and (d). We here investigate the 
band energy 
in the presence  of Zeeman field and SOC interaction while the  sub-lattice mass is absent. 
The edge modes continue [cease]  to exist for  the cut 1 [cut 2] over the strong [weak] bonds even when $|\eta|<0.5$ that is illustrated in Fig.~\ref{fig:2} (a) [(b)] following zigzag ribbon geometries \cite{Wang21}. We demonstrate the energy spectrum of a nano-disc  for cut 2  in Fig.~\ref{fig:2} (c) where the mid-gap states are observed between the  bulk gap carrying the signature of SOTI phase. To be precise,  four mid-gap states live on two corners at $(i,j)=(0,\pm L_y)$  while the remaining two corners at $(i,j)=(\pm L_x,0)$ are empty. This is depicted in the  lower inset of Fig.~\ref{fig:2} (c). The upper inset of Fig.~\ref{fig:2} (c) depicts the four-fold degeneracy of the mid-gap states that is achieved by tuning $g$.  The above SOTI phase is designated by $p_y$ that remains $0.5$ everywhere in $\phi$ except for $\phi=0,\pm \pi$ (see Fig.~\ref{fig:2} (d)). Notice that inside the blue-shaded region, the  spin spectrum gap is non-zero
where 
$p_x$ exhibits monotonic variation. As a result, the Wannier centre lies in the middle of a strong bond safeguarding the corner modes in cut 2 under IS.\\

\indent
A gap in the edge states, traveling on the boundary of the nano-disc in cut 2,  is caused by the breaking of 
the $C_3$ symmetry breaking with $|\eta|<0.5$. Interestingly, 
at $(i,j)=(0,\pm L_y)$, this gap is no longer present. 
As a result, the domain wall is only established at corner  $(i,j)=(0,\pm L_y)$  in this geometry. This  precisely lies on the lattice where the two neighboring weak bond cuts meet precisely at a strong bond \cite{Wang21}. The entire QSHI phase, as displayed in Fig.~\ref{fig:1} (c) and (d), does not transform into the SOTI phase rather the SOTI phase is incubated by the interior part of the above first-order phase. 
In the case of 
the square lattice the 
quadrupolar insulator phase appears out of the entire QSHI phase  under $C_4$ symmetry and TRS breaking \cite{Dumitru19,nag19,schindler-2018}. 
This is markedly different as compared to the present case of honeycomb  lattice. Furthermore, the half-integer quantization in $p_y$ continues to exist without any constraints on the energy of the mid-gap corner modes and the associated degeneracies, see Figs.~\ref{fig:2} (c) and ~\ref{fig:2} (d). \\

\indent
We now  focus on the three terms with $M, V_{\rm so}$ and $g$ to understand the nature of mid-gap corner states. Notice that $g$ and $V_{\rm so}$ break spin degeneracy while $M$ does not. 
The above fact introduces an interesting   spin  polarization  of the mid-gap states. For $g = 0$ and $M = 0$, the finite value of SOC results in positive and negative energy mid-gap states. They exhibit spin polarization and there exists a gap between them.
We refer to these  modes  as spin-polarized pairs.  Such pairs are illustrated for $g \ll V_{\rm so}$ in Fig.~\ref{fig:2} (c) where blue and red represent the spin-up and down, polarization, respectively.
It is therefore noteworthy that 
the gap between these spin-polarized mid-gap states can be manipulated
by varying $g$. Here comes the concept of intra-pair and inter-pair gap for the mid-gap modes due to the complex interplay between the above terms.  In the present case,  inter-pair gap is only renormalized while leaving the intra-pair gap unaltered. Consequently, two pairs of zero-energy corner modes appear in the mid-gap region. 
There exist 
two opposite spin-polarized corner modes, residing in  
each of the pairs, under a suitable choice of parameters.


\subsection{\textcolor{black}{$g=0$ case without inversion symmetry }}

\begin{figure}[!htb]
	\includegraphics[width=0.48\textwidth]{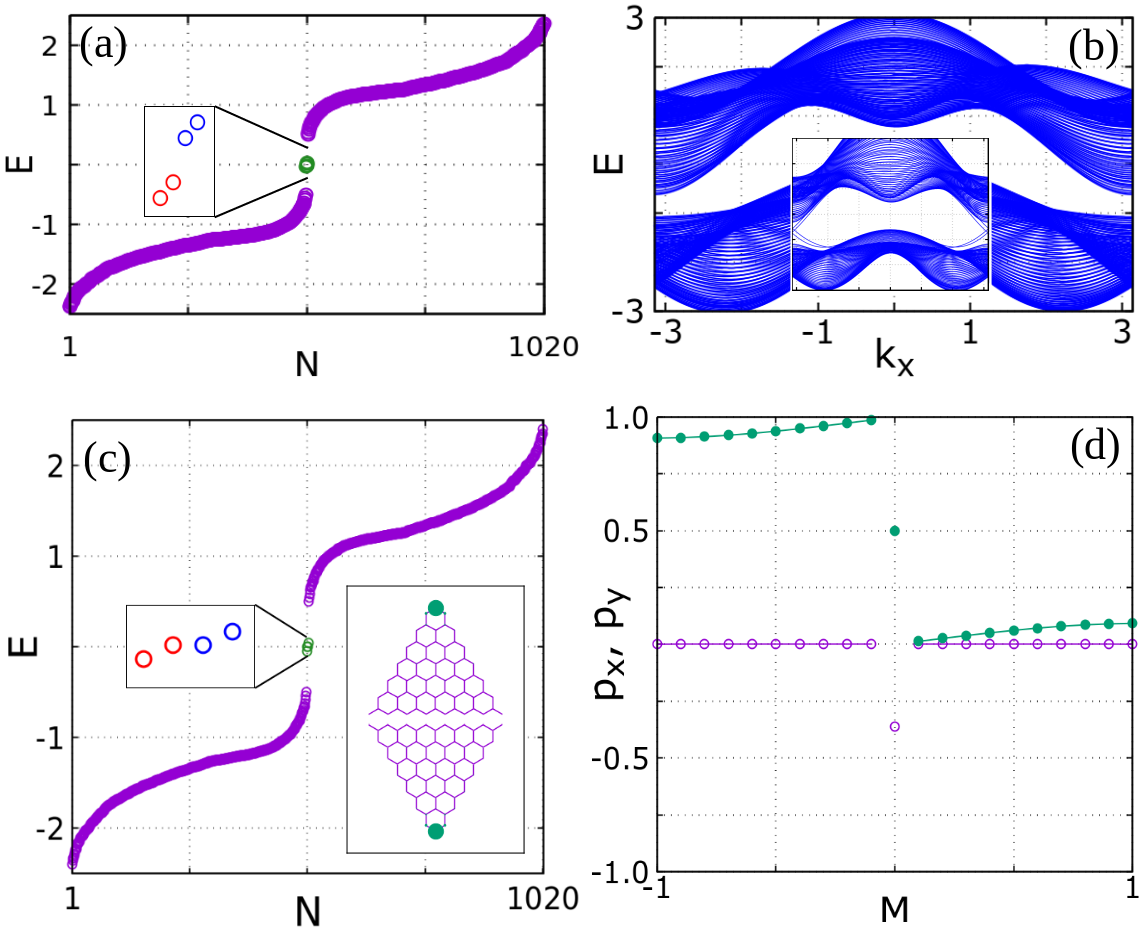}
\caption{Figure (a) shows the
energy bands ($E$ vs $N$) for 
nano-disc with $M= -0.01$ where 
four non-degenerate mid-gap states appear at finite energy.	The  
ribbon 
geometry band structure, considering cut 2, shows 
gap as  depicted in (b) with $M=-0.037$.  The cut 1 band structure shows 
gapless edge states as depicted in the inset of (b).  In (c), we demonstrate the 
nano-disc geometry energy bands for  $M=-0.037$ where doubly degenerate and non-degenerate 
mid-gap corner states appear at zero energy and finite energy, respectively. 
Figure (d) demonstrates a situation where  
even in the presence of zero-energy corner states, the bulk-polarization $p_y$ (green solid circle) deviates from $0.5$ for $|M|\ne 0$. We show the vanishing profile of $p_x$ (pink hollow circle).  In order to make sure that we are inside the SOTI phase as shown in Fig.~\ref{fig:1} (d), the 
parameter considered are $\phi = 0.1$, $V_{\rm SO} = 2.5$, and $g = 0.0$.  After \cite{saha-23}.}
	\label{fig:3}
\end{figure}

We now discuss the SOTI phase appearing for the parameter set $M\neq 0,V_{\rm {so}} \neq 0$, and $g = 0$. In this case, the underlying QSHI phase embeds 
the SOTI phase when $|\eta|<0.5$ as demonstrated in Fig.~\ref{fig:1} (d) \cite{saha-23}. Note that the bulk gap is controlled by $V_{\rm so}$ while the finite sub-lattice mass introduces a degeneracy breaking for the  mid-gap states at non-zero energy. With the variation in  $M$, one enforces 
two out of four mid-gap states 
to appear at zero-energy. An apparent attraction between these two modes occurs.  On the other hand, there will be a repulsion taking place between the 
remaining two modes.  They approach the bulk valence and conduction energy levels while moving further away from the zero energy. As a result, the IS 
breaking yields 
two-fold degeneracy of the corner states at zero-energy while it started from a  four-fold degeneracy.   The four non-degenerate mid-gap states are shown in  Fig.~\ref{fig:3} (a). We demonstrate the repulsion and attraction  among the zero-energy modes as discussed above in 
Fig. \ref{fig:3} (c) where doubly degenerate and non-degenerate corner states appear respectively at zero-energy and 
finite-energy  from the underlying QSHI phase. We further explore the ribbon geometry
band dispersion, considering cut 2, for the above situation in Fig.~\ref{fig:3} (b) that displays a  clear gap. As far as the invariant is concerned, Fig.~\ref{fig:3} (d)
shows that $p_y$ always stays  close to unity (zero) 
for positive (negative) values of $M$.  This 
is an instance where 
the bulk-dipole moment fails to characterize
the corner modes in the SOTI phase.

The spin polarization of these mid-gap pairs can now be examined.  
In this case, positive and negative energy mid-gap state pairs are comprised of spin-up and spin-down states. Extending beyond the 
$M= 0$ case,
the intra-pair gap is induced for a given spin-polarized pair for $M\ne 0$. At the same time, the inter-pair, as well as intra-pair gaps, are controllable by 
$V_{\rm so}$. A competition between $V_{\rm so}$ and $M$ is
found to be responsible for tuning the gaps among such pairs.
One can exemplify a situation of gap tuning by the relative movements of the modes while changing the relevant parameters.   
One mode from a given spin-polarized pair, residing at positive energy,
can move away from its counterpart within the above pair. This is also accompanied by a similar kind of movement in the other spin-polarized pair residing at negative energy. As a result, two modes,  originating from the opposite spin-polarized pairs, eventually approach each other and come closer.  The variation in  $M$ results in the above dynamics where one corner mode from each spin-polarized pair can come arbitrarily close to zero-energy. Now coming to the simultaneous movement of the 
remaining corner modes with opposite spin polarizations, we find that these modes can still reside inside the bulk gap away from zero-energy.
The two-fold degenerate zero-energy mid-gap modes out of four mid-gap modes can appear due to above mentioned interesting evolution of the mid-gap corner. This dynamics thus enables us to
control the 
degree of degeneracy at  zero-energy that  is  evident from Fig.~\ref{fig:3} (c). Consequently, two 
opposite spin polarization profiles are expected from the above pair of zero-energy corner modes.   This is markedly different from 
$M=0$ IS preserved case where 
four-fold degeneracy, coming from two pairs of corner modes, appears at   zero-energy.


\begin{figure}[!htb]
    \includegraphics[width=0.48\textwidth]{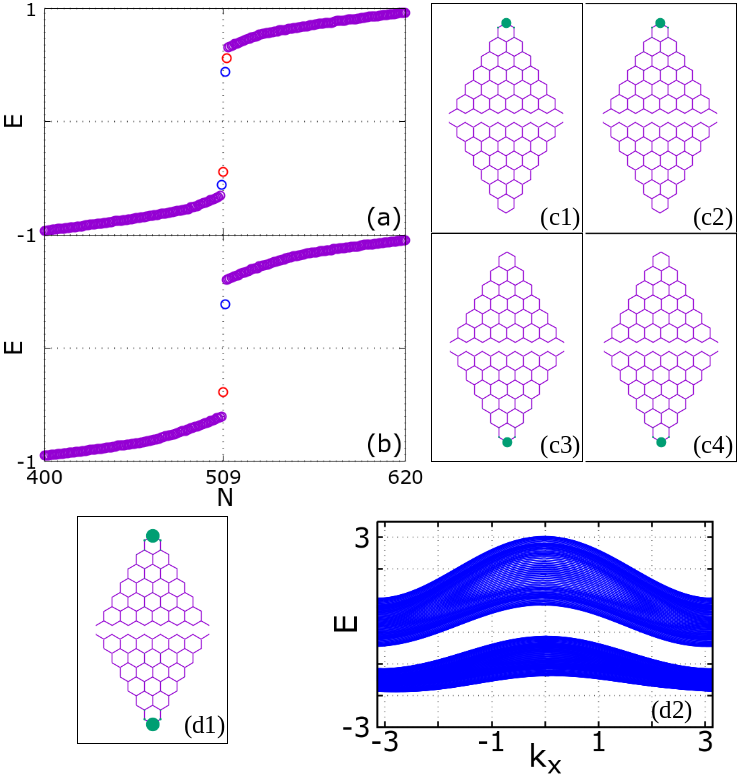}
\caption{Figures (a) and (b) respectively, depict the nano-disc geometry energy bands, adopting cut 2, 
for $g=-0.06$, and $-0.11$. We here consider $M = 0.5$, $\phi = 0.1$, and $ V_{\rm so} = 0.0$. 
In (c1), (c2), (c3) and (c4), respectively, we illustrate the 
spatial localization of the top most red, blue, red, and bottom most blue circles in (a), denoting the four mid-gap
modes. The two mid-gap states at finite-energy found in (b) 
also demonstrate the corner localization.  The  ribbon geometry band structure for (b) is shown in  (d2). After \cite{saha-23}.}
\label{fig:4}
\end{figure}

\begin{figure}[!htb]
\includegraphics[width=0.48\textwidth]{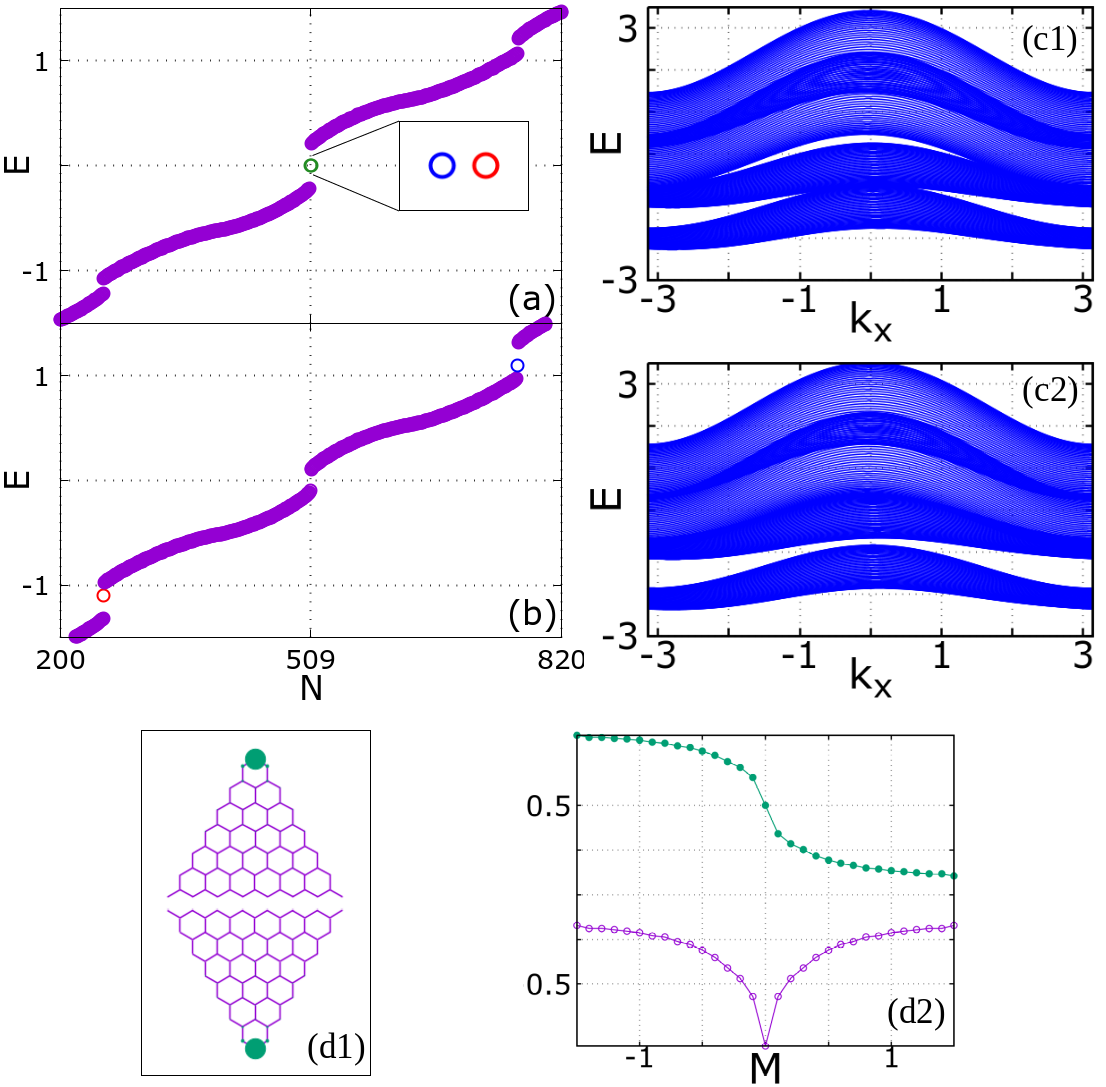}
\caption{We show 
nano-disc energy bands, adopting  cut 2,  for $g=-0.5$ and  $-0.6$ in (a) and (b) respectively. We consider $M = 0.5$, $\phi = 0.1$, and $ V_{\rm so} = 0.0$. In (c1) and (c2), we show the ribbon geometry band structure, corresponding to  (a) and (b),  respectively. 
The two mid-gap states at zero-energy as shown in (a)
are found to be corner-localized in (d1).   In (d2), we show the evolution of $p_x$ (pink hollow circles) and $p_y$ (green solid circles) with $M$ keeping $|M|=|g|$. After \cite{saha-23}.}
	\label{fig:5}
\end{figure}

\subsection{\textcolor{black}{$V_{\rm so}=0$ case without spin-orbit coupling}}


This particular choice refers to two copies of anisotropic IS broken Haldane model with Zeeman field. This is an interesting outcome due to the interplay between $M$ and $g$  while keeping $V_{\rm so}$ fixed at zero. Such a competition between the above two parameters, depending on their relative signs, can push two mid-gap states towards zero-energy while the remaining two mid-gap states move into the bulk bands.
For $|g|<M$,
the variation of mid-gap states in the energy landscape is illustrated in Figs. \ref{fig:4} (a), (b) by varying $g$ and keeping $M$ fixed at $0.5$. Figures. \ref{fig:4} (a),  and (b) are repeated in  Figs. \ref{fig:5} (a), and (b), respectively, for $|g|\ge M$ \cite{saha-23}. 
There exists an 
interesting situation 
for $M = -g$, 
where two mid-gap states out of four are present at zero-energy. Importantly, an increase in $|g|$ causes a reduction of 
the bulk gap around zero-energy.  For $|g|<M$, we notice a gap appearing at finite energy in the 
continuous bulk bands. For $|g|>M$, the mid-gap corner state can be present within the above gap. 
We present 
individual and combined localizations for the mid-gap states, respectively,   in Figs. \ref{fig:4} (c1-c4) and Fig. \ref{fig:4} (d1).  This type of localization is significantly different from the quadrupolar insulator. The  characteristic difference compared to the present case  lies 
in the fact that each of the mid-gap states of  quadrupolar insulator occupies more than a single corner of a 2D square lattice \cite{nag19,Nag21,Ghosh21,Ghosh21b}.  The ribbon geometry
band structure, corresponding to Fig. \ref{fig:4} (b), under cut 2 shows a
gap as observed in  Fig. \ref{fig:4} (d2). This refers to the  SOTI phase as discussed already.
In a similar spirit, 
we find a gap in the ribbon geometry band structure 
as shown in Figs. \ref{fig:5} (c1) and (c2), associated with Figs. \ref{fig:5} (a) and (b), respectively. 
The  topological characterization
in terms of half-integer quantization of $p_y$ is found to be broken for $M\ne 0$ as shown in Fig.~\ref{fig:5} (d2).  \\

\indent
The mid-gap states exhibit 
rich structure in terms of spin polarization that we discuss now by varying $g$ and $M$. To begin with, 
Fig.~\ref{fig:4} (a) demonstrates the opposite
spin-polarization of the finite energy corner modes. The intra-pair energy gap increases  with increasing $|g|$. One of the mid-gap states from each pair eventually disappear into the bulk bands. On the other hand, the remaining mid-gap states have opposite spin-polarization.
Importantly, 
for $|g|=|M|$, these  corner modes, emerging from two different mid-gap pairs, live at  zero-energy. When comparing with 
earlier $g=0$ case, we also find  one pair with opposite spin polarization. However, there exists a basic difference in term of the  number of  mid-gap corner modes. For example, this number reduces to two  from four as obtained for $g=0$ (see Figs.~\ref{fig:3} (a), (c), Fig.~\ref{fig:4} (b) and Fig.~\ref{fig:5} (a)).


\subsection{\textcolor{black}{General case with $M\ne 0$, $g\ne 0$, and $V_{\rm so}\ne 0$}}
	
\begin{figure}[!htb]
\includegraphics[width=0.48\textwidth]{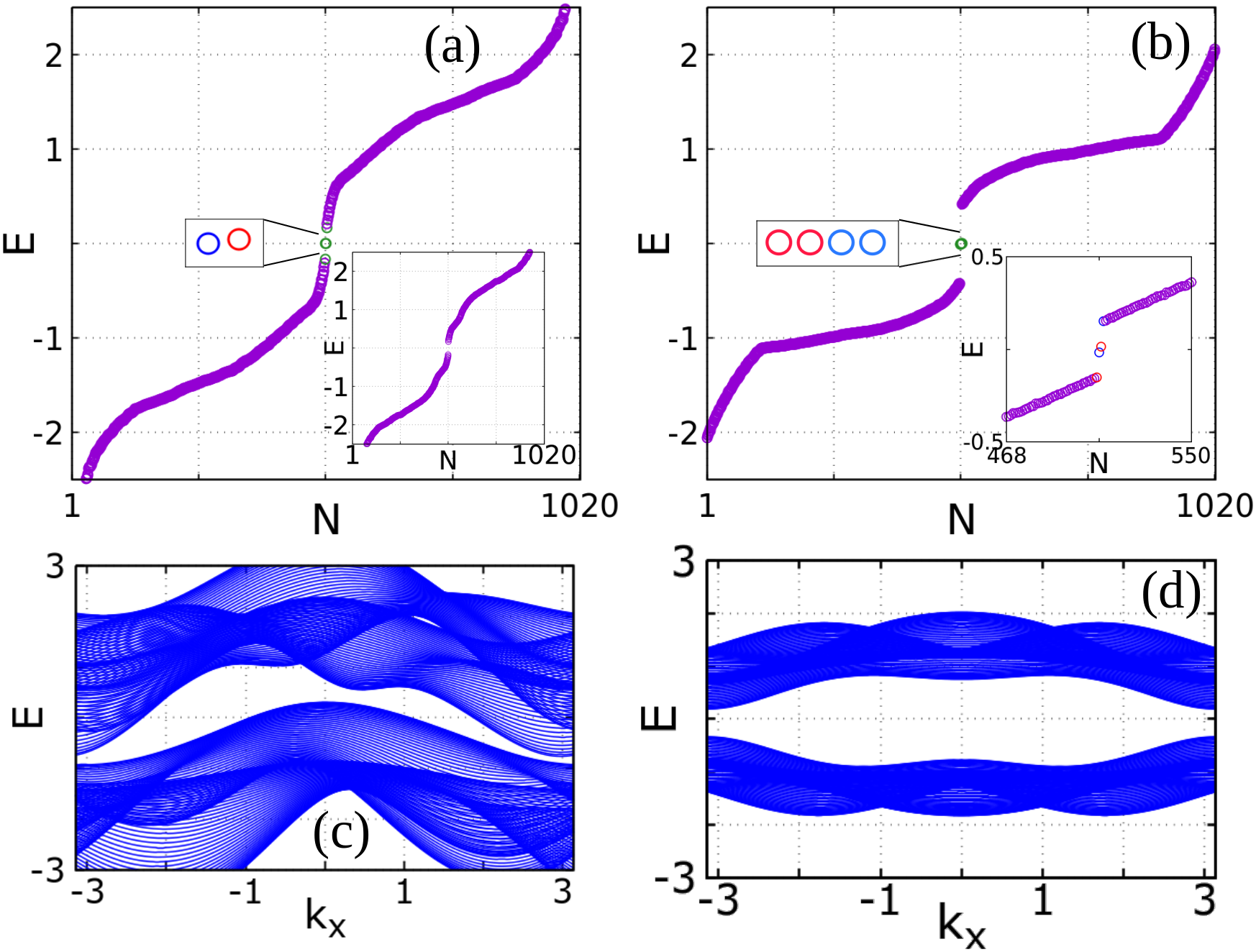}
\caption{
The mid-gap corner modes, considering  $(M,V_{\rm so},g,\phi,\eta)=(-0.6,2.5,-0.33,-0.1,0.25)$  and $(0.0, 0.5, 0.074,1.5, 0.25)$ are demonstrated in (a), and (b), respectively. We make sure that the SOTI phases (a) and (b) are taken from the underlying QSHI, and QAHI phases as displayed in Fig.~\ref{fig:1} (b) and (c), respectively.
Inset (a) shows an instant where there exist no the mid-gap states in the energy dispersion  for  $(1.2,2.5,-0.33,-0.1,0.25)$. Inset (b) displays a situation where there exist two mid-gap states 
in
the energy dispersion  for $(-0.4,0.5, -0.33, -1.57,0.25)$ that is taken from the underlying QAHI phase in Fig.~\ref{fig:1} (a). 
The ribbon geometry band structure is gapped out as illustrated in (c) and (d) corresponding to (a) and (b), respectively.  After \cite{saha-23}.
}
	\label{fig:6}
\end{figure}

We now examine the mid-gap corner modes in the SOTI
phase for 
a generic set of parameters. For this, we demonstrate  nano-disc and ribbon geometry band structures, respectively,  in Figs.~  \ref{fig:6} (a) and (c) \cite{saha-23}. Our study finds that the underlying QSHI phase serves as a precursor 
for  as shown in Fig.~  \ref{fig:1} (b).
This causes the SOTI phase to become encapsulated within the first-order QSHI phase. However, 
QSHI phase as a whole does not morph into the   SOTI  phase rather 
the SOTI  phase is only embedded in the interior part of it for $M,g\ll V_{\rm so}$, $\phi < \pi/4$. Such phase boundaries can be qualitatively understood from the effective band gaps for $|\eta| < 0.5$ and are depicted as an inset in  Fig.~  \ref{fig:6} (a).
Once  we chose the parameters
from an exterior part of the underlying QSHI phase, the 
mid-gap corner states are absent. 
A similar tendency is noticed in Fig.~\ref{fig:1} (c) and (d). However, the phase diagram of the SOTI phase in $M-\phi-g-V_{\rm so}$ parameter space, demarcating the exact boundaries, requires further studies.
This is also due to the fact that the appropriate topological invariant is yet to be determined for the IS broken case. The analysis is now extended to QAHI phase as shown in Fig.~~\ref{fig:1} (c) with $V_{\rm so}=0.5$ and $M=0$. We find four mid-gap corner modes at zero-energy 
when the  SOTI phase preserves the IS. Such a phase emerges from the underlying QAHI phase when $\eta=0.25$, see Fig.~ \ref{fig:6} (b). We examine the 
ribbon geometry band structure
in Fig.~ \ref{fig:6} (d)
where 
the absence of the edge modes is 
suggestive for the  SOTI phase. We extend this to the case for $V_{\rm so}=0.5$ and $M\ne 0$ from Fig.~\ref{fig:1} (a) where
the signature of inversion broken SOTI phase, embedded in  the underlying QAHI phase, is manifested into 
two mid-gap corner modes,  see the inset in Fig.~ \ref{fig:6} (b). 



\indent

As discussed earlier SOTI phases emerge for $0<\eta<0.5$. To understand this, it is useful to examine the band energy of FOTI phases for $\eta=0.5$. Hamiltonians at the ${\bm M}$-points are gapped out as a prerequisite for the SOTI phase to appear. 
The energies for the spin up and spin down sub-blocks  at ${\bm M}$ points are given by,
\begin{eqnarray}
E^{\pm}_{\uparrow}=-V_{\rm so} \sin \phi/3  + g  \pm  \sqrt{(2\eta t_{1} - t_{1})^2+M^2}&& \\
E^{\pm}_{\downarrow}=V_{\rm so} \sin \phi/3  - g  \pm  \sqrt{(2\eta t_{1} - t_{1})^2+M^2} .&&
\end{eqnarray}
For $\eta <0.5$ and even for $M=0$, irrespective of the values of $V_{\rm so}$ and $g$,
the diagonal blocks are always individually gapped out. The complete $4$-level Hamiltonian, interestingly, is also gapped out for $g=V_{\rm so} \sin \phi/3$. 
Coming to the energetics of the Hamiltonian, the energy associated with spin-up and spin-down sectors match each other.
One can encounter a situation for $M=-g\ne 0$ and $V_{\rm so}=0$ where not only individual spin sectors are gapped out but also  two different spin sectors are energetically separated.
A qualitatively identical gap structure of the energy bands are observed for $M,V_{\rm so}\ne0$, and $g=0$ case as well. It helps us to understand the previous finding 
where $g$, $\phi$ and $V_{\rm so}$ 
can be tuned to 
manipulate
the energy separation between the spin up and down mid-gap corner modes. On the other hand, all spin sectors are identically gapped by $M$.  Altogether it represents a fairly complicated interplay that is  clearly noticed in Figs.~\ref{fig:2}, \ref{fig:3}, \ref{fig:4}, \ref{fig:5} illustrating the emergence of even or odd number of pairs of zero-energy modes under suitable parameters. It can also happen  in  several occasions that any two out of the four energy levels $E^{\pm}_{\uparrow,\downarrow}$ become degenerate.
In this case, Hamiltonian at ${\bf M}$-point no longer remains gapped out resulting in a collapse of the SOTI phase. \\

\indent
The FOTI phase continues to exist for $0.5\le\eta\le 1.0$ where the gapless Dirac points ${\bm K}^{\eta}_{\pm}$ shift from  the ${\bm K}$-points as obtained for $\eta=1$.  The gap at the Dirac points ${\bm K}^{\eta}_{\pm}$ for spin-up and spin-down sectors are found to be $\Delta E^{\pm}_{\uparrow}=n^{\pm}_4-n^{\pm}_3$ and  $\Delta E^{\pm}_{\downarrow}=n^{\pm}_4+n^{\pm}_3$ (neglecting the $n^{\pm}_5$ term that is associated with identity for a given spin sector) when the Hamiltonian is written in the block-diagonal form in the basis $(c_{A\uparrow},c_{B\uparrow},c_{A\downarrow},c_{B\downarrow})$.  For the FOTI phases to exist  $\Delta E^{\pm}_{\uparrow (\downarrow)} \Delta E^{\mp}_{\uparrow (\downarrow)} <0$.  However, such a criterion is also expected to hold  for SOTI phases that are embedded in the underlying FOTI phases with both spin channels being non-trivial. However, the spin degree of freedom is intertwined with sub-lattice leading to much more complicated constraints for SOTI phases. As a result,  the SOTI phases are hard to understand from  the underlying $2$-level  spin polarized Hamiltonian.

\section{Emergent or extended symmetries and their consequences}


The modern theory of classification of topological systems is determined by the symmetries it possesses. One may note that a tight-binding model on a given lattice can be represented by a Hermitian  matrix. Except for this basic property, the Hamiltonian can possess other symmetries which eventually decide its topological classifications. Based on chiral symmetry $\mathcal{C}$, particle-hole symmetry $\mathcal{P}$ and TRS $\mathcal{T}$ the 10-fold classification was obtained for the topological matter \cite{10fold}. It may be noted that these symmetries do not refer to the lattices or the space-groups in which the model is defined but they are \textcolor{black}{emergent/extended} symmetries intrinsic to the nature of the model.  In this regard, for translational invariant systems, the relation of energy eigenvalues and wavefunctions at different $k$-points are important and governed by the physical processes such as hopping defined on the lattice. The chiral symmetry $\mathcal{C}$ refers to the existence of a unitary transformation which anti-commutes with the Hamiltonian as $\mathcal{C} H \mathcal{C}^{-1}=-H$ such that $\mathcal{C} \mathcal{C}^{\dagger}=\mathcal{C}^2=1$. In the case of  the particle-hole (or charge-conjugation) symmetry,  one refers to the existence of an anti-unitary and anti-commuting operation such that $\mathcal{P}H\mathcal{P=-H}$ with $\mathcal{P}\mathcal{P}^{\dagger}=1, \mathcal{P}^2=\pm 1$. Finally the TRS $\mathcal{T}$ requires an anti-unitary operation commuting with the Hamiltonian as $\mathcal{T} H \mathcal{T}^{-1}=H$ with $\mathcal{T}^2=\pm 1$. The chiral symmetry is generally associated with the handedness of a physical process.  For multicomponent wavefunction it represents the change of certain emergent vector quantity as we change a suitable parameter in the Hamiltonian \cite{felser-2022,galitski-2021}. On the other hand, in band insulator particle-hole symmetry reflects the invariance of net energy for a $N-1$ electron system with that of $N+1$ electrons. We note that here usually we consider  $N$ electron system  as a reference state \cite{ph-zirnbauer}.

\textcolor{black}{We here note that we deal with non-interacting systems only and hence the word 'emergent' or 'extended' suggests the symmetries that are possessed by the single-particle Hamiltonian in addition to the spatial and ${\mathcal P}, {\mathcal C}, {\mathcal T}$ symmetries. We further note that the emergent symmetries should not be confused with the symmetries of an effective non-interacting model derived from an initial interacting model that does not have the emergent symmetries due to collective effects.  In the present case, the symmetry relation associated with  the emergent symmetry operator requires a parameter of the non-interacting Hamiltonian yielding an extended parameter space to encompass that emergent symmetry. In other words, the emergent symmetry imposes a constraint on the choice of the parameter namely, magnetic flux in the  model, and hence an extended parameter space suffices the symmetry relation to be satisfied. As a result, the emergent symmetry can also be considered as extended symmetry of the non-interacting Hamiltonian which is why we use the words emergent and extended interchangably. } \\

\indent
We reiterate that the Haldane model~\cite{haldane-1988} does not preserve TRS the Kane-Mele model\cite{kane-2005-1st,kane-2005-2nd} does. For the present model which has been defined on Kane-Mele model by incorporating staggered flux distribution of Haldane model, the TRS is no longer preserved. As the model shows intriguing topological phases and the associated edge states we now elucidate the \textcolor{black}{emergent/extended} symmetries of the edge modes for the model Hamiltonian given in Eq. (\ref{main:ham1}). A look back into Figs.~\ref{fig:edgeStateVsoGTVrfoti} (g) and (h) would be helpful to remind ourselves about the structure of edge states. In QAHI phase, with $\phi=\pm \pi/2$, one has zero energy chiral edge states. For other values of $\phi$, one encounters finite energy edge states for a given topological phase. The avoided level crossing structure, as found in  QSHI phase with magnetic doping  and exchange field  \cite{sheng-2011,Liu-2008,li2013}, are not observed for the edge modes in the present case.  This  suggests that there is a qualitative difference  as far as the behaviour of edge state is concerned. It may indicate that the way one breaks TRS is important and hence the staggered magnetic field produces qualitatively different edge states than that of magnetic doping and exchange field.  For the present case though TRS is broken due to staggered magnetic field, the energy dispersion depends on $\phi$ and $k$ as follows $E(\pi-k, \phi)= E(\pi+ { k}, - \phi)$. This hints that an implicit TRS is established for the system considered here in general.
 Further we find that $E(\pi -{ k},\pi -\phi)=-E(\pi+{k},\pi + \phi)$ following a composite anti-unitary symmetry. These two symmetry protects the edge modes.  Therefore, the two-fold effect of these anti-unitary symmetries allows us to map $\rm C_{\sigma} \to - \rm C_{\bar{\sigma}}$ where $\sigma$ and $\bar{\sigma}$ denote spin indices with $\sigma \ne \bar{\sigma}$. Once we combine the effect of implicit TRS and anti-unitary symmetry, the chiral edge modes are guaranteed to have zero energy at   $\phi= \pm \pi/2$. \\

\indent
As far as the bulk Hamiltonian is concerned, a careful analysis suggests that there exists an  \textcolor{black}{emergent/extended} symmetry operator, generated by ${\mathcal B}=\sigma_2 \tau_1$, such that  ${\mathcal B} H(k_x,k_y,\phi){\mathcal B}^{-1}= -H(k_x,-k_y,\pi-\phi)$. The symmetry operator maps $A\uparrow \to B\downarrow$ and $ A\downarrow \to B\uparrow$ \cite{Saha21, saha-23}. Note that both the spin-sectors are topological in QAHI $C_{\uparrow}=C_{\downarrow}\ne 0$ and QSHI $C_{\uparrow}=-C_{\downarrow}\ne 0$ phases as extensively demonstrated in  Ref.\cite{Saha21} with  $C_{\uparrow}$ ($C_{\downarrow}$) representing the spin-Chern number for $\uparrow$ ($\downarrow$). Only one spin sector is topological for QASHI such that  $C_{\uparrow}=0,C_{\downarrow}\ne 0$ or $C_{\uparrow}\ne 0,C_{\downarrow}= 0$. This topological character of the  underlying first-order phase is very important in order to embed the emerging HOTI phases under suitable conditions.  To be precise, both the spin sectors in the FOTI phases have to be topological in nature to host the HOTI phases. The symmetry operator interchanges the spin degrees of freedom. Therefore, once both the spin sectors have non-trivial topology, the above symmetry operation is only permitted causing the QAHI and QSHI phases to give rise to HOTI phases. Such symmetry operation does not permit the QASHI to host the HOTI phases as only one of the spin sectors is topological. This symmetry helps us to predict the formation of SOTI phases in the phase diagram plotted with respect to staggered magnetic flux $\phi$. Note that the present system always preserves the \textcolor{black}{emergent/extended} symmetry irrespective of the values of $\eta$.  Therefore, the above analysis can also satisfactorily explain the absence of QASHI phase for IS preserved case with $M=0$ (see Figs.~\ref{fig:1} (c), (d)). Hence SOTI phase can not emerge from the underlying QASHI phase that is itself absent for $M=0$.

\textcolor{black}{It is pertinent to mention the role of space group symmetry such as rotational symmetry  ($C_3$) and inversion symmetry (${\mathcal I}$). $C_3$ refers to the $120^{\circ}$ rotational symmetry in graphene and it also holds true in case of original Haldane model as well as the extended Haldane model given in Eq. (\ref{main:ham_Haldane}). Adding spin-orbit interaction which considers equal amplitude of the hopping along three bonds also preserves the $C_3$ symmetry. One can indeed easily show that $C_3 H C^{-1}_3=H$. Coming to   the inversion symmetry ${\mathcal I}$ for honeycomb lattice refers to the inversion of the lattice with respect to the mid point of a $Y$-like bond. Under this process,  $A$-sublattice is mapped to $B$-sublattice. We note that the original Haldane model does not preserve it due to the presence of mass term `$M$'. We note that in honeycomb lattice inversion symmetry is connected to chiral symmetry. It is important to understand the distinctive  role played by particle-hole symmetry ($\mathcal{P}$), the \textcolor{black}{emergent/extended} symmetry operator ($\mathcal{B}$), the time-reversal symmetry ($\mathcal{T}$), $C_3$ symmetry and inversion symmetry $(\mathcal{I})$ in governing the FOTI and HOTI physics. The FOTI phase appears in the presence of $C_3$ symmetry and continues sustaining even after  breaking $C_3$ symmetry. The inversion symmetry $\mathcal{I}$ was essential in obtaining the FOTI phase in Haldane model though it is not absolutely necessary when spin-orbit interaction was present. On the other hand, for the SOTI phase, the $C_3$ symmetry breaking and presence of inversion symmetry ($\mathcal{I}$) is a necessity. However, interestingly, there is region where localized zero-dimensional corner modes appear without  a quantized dipole moment  in the absence of $C_3$ and $\mathcal{I}$ symmetries. Whether this particular charactaristic is topological or not is to be reexamined and we hope that the \textcolor{black}{emergent/extended} symmetry  operator $(\mathcal{B})$ will be useful to provide such characterization in the future.}

For IS broken case with $M\ne 0$, when QASHI  is present in the phase diagram, the SOTI phase can in principle emerge that we leave for future studies. 
In principle, SOTI phase can emerge from QASHI phases with non-zero mass ($M \ne 0$). However, we leave this for future studies. Interestingly when the sublattice mass ($M$) and the Zeeman field strength $g$ are equal in magnitude but opposite in sign and $V_{\rm so}=0$, one finds  from the Hamiltonian (Eq.~(\ref{main:ham})) that the terms $c_{A\uparrow}^{\dagger} c_{A\uparrow}$ and $c_{B\downarrow}^{\dagger} c_{B\downarrow}$ (and $c_{A\downarrow}^{\dagger} c_{A\downarrow}$ and $c_{B\uparrow}^{\dagger} c_{B\uparrow}$) appears identically.  It establishes an interesting reciprocity where the role of spin and sublattice degrees of freedom are intertwined as evident from $\mathcal{C}$. As a consequence, each spin sectors seem to yield a mid gap state with zero energy as shown in Fig.~\ref{fig:5} (a), which is remarkable because IS is broken in this instance. That being the case one may conclude that this \textcolor{black}{emergent/extended} symmetry determines the number of zero energy mid gap states and their spin-polarization. For details refer to Figs.~\ref{fig:2} and ~\ref{fig:3} where we demonstrate the  existence of  even and odd pairs of zero-
energy mid-gap states with opposite spin polarizations.


We now discuss the consequences of this \textcolor{black}{emergent/extended} symmetries on the dipole moment $p_x, p_y$ calculated for the model Hamiltonian given in Eq. (\ref{main:ham1}). The model breaks IS due to finite $M$ and it is found in earlier study\cite{Wang21} that anisotropic Haldane model does not yield a quantization of $p_y$ for finite $M$ though the mid-gap states survive (not at zero energy). The quantization of dipole-moment and existence of mid-gap zero energy corner modes are results of Chiral symmetry as discussed at great length previously \cite{benalcazar-2017-prb}. Here we would like to discuss the characteristics features associated with the \textcolor{black}{emergent/ extended} symmetry given by ${\mathcal B}=\sigma_2 \tau_1$, such that  ${\mathcal B} H(k_x,k_y,\phi){\mathcal B}^{-1}= -H(k_x,-k_y,\pi-\phi)$. Lets consider a super-partner of the Hamiltonian $H(k_x,k_y,\phi)$ defined by $H(-k_x,k_y,\pi-\phi)$ while the composite Hamiltonian is defined as $\mathcal{H}(k_x,k_y)=H(k_x,k_y,\phi)\otimes \mathbbm{1} + \mathbbm{1}\otimes H(-k_x,k_y, \pi-\phi) $. Here $\mathbbm{1}$ denotes a $4\times 4$ unit matrix. The Hamiltonian $H(-k_x,k_y, \pi-\phi)$ can be thought of as another Haldane model with $\phi \rightarrow \pi-\phi$ and $V_{\rm so}, V_R$ and $t_2$ changing sign as evident from Eqs. (\ref{eqn01}) to (\ref{eqn9}) and from  definition of $f,g,h$ as given in Eqs. (\ref{eqg}) to  (\ref{eqh}). Now if we construct a super chiral operator defined as $\mathcal{C}_s= \mathcal{C}\otimes \mathbbm{1} + \mathbbm{1}\otimes \mathcal{C}$. With this defnition of $\mathcal{C}_s$, it is easy to check that ${\mathcal C}_s \mathcal{H}(k_x,k_y,\phi){\mathcal C}_s^{-1}= -\mathcal{H}(-k_x,-k_y,\phi)$. Thus the dipole moment defined for such extended Hamiltonian should exhibit quantization. The realization of such extended Hamiltonian in a tight binding models is  a scope for future study.

\section{Discussion and future perspective}

As mentioned in the beginning, the purpose of the review is to illustrate the role of different parameters within the scope of tight binding model.  The specific questions that we focus on are the following:  how different topological phases arise, what are the appropriate bulk topological invariants that characterise these phases, how are the gap profiles change for such topological phase transitions in various cases, and finally how do the boundary modes carry the Hallmark signature of a given topological phase. These questions constitute the central themes of the topological aspects of any solid state system irrespective of the details of the underlying theoretical framework. There exists theoretical model building in one, two and three dimensions for respective FOTIs and HOTIs \cite{armitage-2018-rmp,QHE_review3,Ching-kai-2016-rmp,wen-2017-rmp,ludwig-2017-scripta,zhang-2011-rmp}, as well as experimental realizations of these models are also made in acoustic system\cite{ma-chan-nature-review}, mechanical systems\cite{zheng-2022-review}, phononic, photonic\cite{ozawa-2019-rmp,zhihao-2022}, meta-materials\cite{rocklin-2017,xiang-2022} and  magnetic systems\cite{tokura-2019,bernevig-nature-review-2022}. In each of these systems, there is an underlying challenge and understanding these barriers for simple theoretical models always brings in insightful correlations between symmetry, boundary states and topological invariants. With the above broad perspective in mind, it is essential to summarize the various interesting outcomes that an extended Haldane model offers to us. This unifies many aspects of FOTIs and SOTIs within a single model and yet gives rise to engaging questions that are to be investigated in the future studies. \\

Regarding the aspect of FOTI phases, we considered a parametric enlargement of Haldane model to spinful case which takes into account spin-orbit couple and also signature staggered magnetic field. This model can also be thought of as an useful adaptation of Kane-Mele model into Haldane model. \cite{Saha21}.The outcome of such a consideration is seen remarkable with the presence of many new topological phases such as  QASHI, QAHI and QSHI phases. There is also an extended critical region whose boundary is determined by multiple topological phases. This extended critical region serves as an interesting precursor region and how different topological phase transitions are governed by certain underlying hidden order parameters that need further investigation. All these phases mentioned are characterized by the finite spin-Chern number $(\rm{C}_{\sigma}$ for a given component $\sigma=\uparrow, \downarrow$.  The QASHI phase, characterized by $(0,\rm{C}_{\downarrow})$ and $(\rm{C}_{\uparrow},0)$. This imply that the spin-component with vanishing topological index is trivially gapped out and the spin component with finite $\rm{C}{\sigma}$ has topological gap. As a result we have chiral edge charge current which is polarized with the spin-component $\sigma$  i.e it is  spin-selective. 
In the QSHI and QAHI phase the spin-Chern index for both the spin-component is non-zero. While for QSHI it is opposite to each other, for QAHI it is identical. This means that for QSHI, the charge edge current at a given edge cancels each other due to subtraction from both the the channel. However the net spin-edge current remains finite hence it corresponds to  spin-polarized but charge neutral current. Finally for QAHI phase the edge current from both the spin channel add up doubling the edge current. However the net spin-transport cancels due to the opposite contribution from each spin-channel and hence constitute a spin-neutral charge current. This intriguing spin polarization of the edge modes could be very interesting as far as quantum technology aiming to exploit robustness of the topologically protected edge modes are concerned. \\

 After the theoretical progress being  made to understand topological aspect of various models, there has been intense effort to realize such models in feasible experiments. As mentioned, one of the key ingredients of topological insulator is spin-orbit coupling and in real material its strength is intrinsically determined limiting ones ability to control it and thus realize topological phase transition.  In this aspect  optical lattice platform could serve as a promising area where SOC is theoretically proposed \cite{soc_theory1,soc_theory2,soc_theory3,soc_theory4,soc_theory5} and experimentally realized \cite{soc_exp1,soc_exp3,soc_exp4,soc_exp5}. We note that spin-orbit coupling needs a complex second-neighbour hopping such that its phase depends on particular spin component. In a related study \onlinecite{exp1}, the complex second nearest neighbour hopping which depends on the spin are realized in the presence of a spin-dependent force due to an oscillatory magnetic gradient.

It is always a commendable feat to realize such topological phases in real materials.  To obtain QAHE quantum well system based on HgTe has been used after doping with magnetic ion \cite{Liu-2008}.  Similarly first-principle based calculation speculates that marious if one implants certain transition metal elements (such as Cr, Fe) into  tetradymite semiconductor ( $\rm Sb_2Te, Bi_2Se_3, Bi_2Te_3$ )can results into magnetically ordered insulators which may give rise to QAHE\cite{yu-2010}.  Alternative proposal based on double-layer transition metal oxides with perovskite structure \cite{xiao-2011,cook-2014} and engineered graphene \cite{qiao10,zhang-2012}.offers as candidate material for QAHE.
Remarkably $\rm{\tilde{A}Fe(PO_4)_2(\tilde{A}= K, Cs, Ba,  La)}$ shows a complex second nearest hopping \cite{hskim-2017} (mimicking presence of staggered magnetic field) yielding Chern bands.
All these examples show that the model proposed in this article may be realized if a spin selective hopping can be established in prospective materials like $\rm{\tilde{A}Fe(PO_4)_2}$, or magnetic impurity planted on graphene~\cite{qiao10,qioa-2014} or in transition metal oxide heterosctructures~\cite{xiao-2011}. Such spin dependent hopping can be established by proximity effect from suitable magnetically ordered two-dimensional materials or substrate~\cite{chang-2019,ramon-2020}  or skyrmion lattice \cite{yang-2014,muhlbauer-2011,kurumaji-2019,choi-2021,gilbert-2015} or magnetic insulators such as $\rm{MnTe, MnSe}$. The materials with intrinsic SOC showing QSHE~\cite{kim16,garrity-2013} can turn to be useful candidate materials  if we are able to manipulate the spin-orbit interaction as explained before. In this regard  the external pressure may serve as an alternative way to control the Rashba SOC~\cite{huang-2020} where as the intrinsic SOC can be controlled by suitable doping or adatoms~\cite{kandemir-2013,luis-2015,tobias-2017} by introducing long range coulomb interactions in the system.  In terms of the future applications, our study can also serve as an interesting platform to investigate the the spin entanglement in various phases ~\cite{soc_entanglement,2d_material}. Further using magnetic and non-magnetic disorder, it may be useful to investigate the fate of various first order and second order topological phases  \cite{soc_anderson}. \\

\indent
It is intriguing that different topological phases discussed here are protected due to emergent TRS as well as a composite particle-hole symmetry. The role of spectral band gap and spin-spectrum gap both play crucial roles in determining the topological characterization and stability respectively. Furthermore, the spin-spectrum gap is necessarily  be finite for any topological phase either be of first or second order. We may noe that though spin-spectrum gap for second order phases are to be understood as  spin-spectrum gap, obtained for $|\eta|<0.5$ and  may be called  as second order spin-spectrum gap. To summarise, for first order topological phase finite spin-spectrum gap ensures the edge modes while for the second order topological modes it establishes the appearances of mid-gap zero energy corner modes.  Also for the quantization of the dipole moment $p_y$ the spin-spectrum gap is to be finite as well in the case of IS being preserved.  Interestingly, here we found a scenario such that $p_y$ remains quantized but the zero energy corner modes becomes finite and extending the scope of SOTI physics perspective.  This suggests that the bulk-boundary correspondence for SOTI phases are to be re-examined such that the half-integer quantization of $p_y$ and zero energy mid gap corner modes are not in one to one correspondence always. This probably hints at unknown topological phases whose bulk-boundary correspondences are yet to be understood.  
\indent

Unlike the square lattice with $C_4$ symmetry and the corresponding HOTI phases in $C_4$ symmetry broken system\cite{schindler-2018,benalcazar-2017,benalcazar-2017-prb,parames-2017}, the  HOTI phases in $C_3$ symmetry broken  system \cite{Wang21,Lee20,Ezawa18,Saha21} are found to be more intriguing. This is due to the fact that the appearances of corner modes and also the higher order moments such as dipole-moment need a careful understanding in terms of the  anisotropy present in the particular model of finite size geometry as well as its relation with particular momentum associated with quantized dipole moment.  The existence of finite $p_y$ in the absence of mid-gap states probably reinforces such an understanding.   Here we have examined the evolution of eight different quantum hall phases found  \cite{Saha21} for the extended Haldane model in the  case where hopping are no longer isotropic. Moreover,  the present review shows extensively the effect of multiple parameters such as different masses depending on sub-lattices and Zeeman field and SOC in engineering the corner modes. The open issue that remains unresolved is the criteria for the above SOTI phase to emerge out of the parent  QSHI  and QAHI phases only while the remaining QASHI phases do not have any SOTI analogue as evident for the IS preserved case.  Finaly we comment that the study in topological state of matter and topological insulator in particular is not only driven by the theoretical curiosity but also for practical applications because of its unique properties such as bulk-boundary correspondence of bulk gap and gapless edge modes. Thus it is natural that a system which offers great controlling ability is of high importance. In this regard the system considered here is remarkable in the sense that one can control the number of zero energy corner modes in SOTI phase or the number of one dimensional edge modes in FOTI phases by tuning the model parameters.  
One may note that laser assisted spin-orbit interaction for Bose-Einstein condensate of $Rb$-atoms \cite{soc_exp1}has been achieved and a phase transition was obtained by varying the resulting SOC. For the fermionic counterpart similar laser assisted SOC for $^{40}K$ fermi gasses\cite{soc_exp3} has been  achieved and it is found that the Dirac cones can be controlled. Further in a significant study \cite{soc_exp4}, A perpendicular Zeeman field and SOC have been obtained in the same $^{40}K$ fermi gasses. As the experimental realization of Haldane model is already achieved \cite{exp1,liu2018generalized}, we hope that some of the interesting result discussed in this review offers an interesting platform to verify itself.


{\it Acknowledgement}: S.M. acknowledges  SAMKHYA: High Performance Computing Facility provided by Institute of Physics, Bhubaneswar. T.N. acknowledges the NFSG ``NFSG/HYD/2023/H0911" from BITS Pilani.


\begin{thebibliography}{100}
\bibitem{spin-gap} The spin gap is calculated from the spin projected Hamiltonian
$H_{ij}(\bm k)=\langle \psi_i({\bm k}) | \sigma_0 \tau_3| \psi_i({\bm k}) \rangle $  where 
$|\psi_i({\bm k})\rangle$ denote the valence band wave-functions $i=1,2$.
\bibitem{spin-chern-number} First one obtain two valance band eigenvectors (say $\Ket{V_1}$, $\Ket{V_2}$) from Eq. \ref{main:hamk}. Following this one construct the projector on the valance band as $P = \Ket{V_1}\Bra{V_1} + \Ket{V_2}\Bra{V_2}$. Now we diagonalize the projected spin operator $\tilde{S} = PSP$ and find the eigenvectors of non-zero eigenvalues. Spin-Chern number is defined as the difference of Chern number of these two latest eigenvectors.
\end{thebibliography}
\end{document}